\documentclass[journal,10pt]{IEEEtran}
\makeatletter
\def\endthebibliography{%
    \def\@noitemerr{\@latex@warning{Empty `thebibliography' environment}}%
    \endlist
}
\makeatother

\usepackage{cite}

\usepackage[pdftex]{graphicx}
\usepackage[caption=false,font=footnotesize]{subfig}
\usepackage{tikz}
\usetikzlibrary{arrows.meta, positioning, shapes.geometric}
\usepackage{amsmath}
\usepackage{mathtools, cuted}
\usepackage{amssymb}
\usepackage{bm}
\usepackage{mathrsfs}
\usepackage{breqn}
\usepackage{bbm}
\usepackage{booktabs}

\usepackage{pifont}
\usepackage{xcolor}
\usepackage{url}
\usepackage{lettrine}
\usepackage{lipsum}
\usepackage{siunitx}
\usepackage{soul}
\usepackage{array}
\usepackage[inline]{enumitem}
\usepackage{epsfig}
\usepackage{filecontents}

\usepackage{algpseudocode}
\usepackage{algorithm, tabularx}
\usepackage{multirow}
\newcolumntype{L}[1]{>{\raggedright\let\newline\\\arraybackslash\hspace{0pt}}m{#1}}
\newcolumntype{C}[1]{>{\centering\let\newline\\\arraybackslash\hspace{0pt}}m{#1}}
\newcolumntype{R}[1]{>{\raggedleft\let\newline\\\arraybackslash\hspace{0pt}}m{#1}}
\newlength{\maxwidth}

\makeatletter
\newcommand{\multiline}[1]{%
	\begin{tabularx}{\dimexpr\linewidth-\ALG@thistlm}[t]{@{}X@{}}
		#1
	\end{tabularx}
}
\makeatother
\algdef{SE}[SUBALG]{Indent}{EndIndent}{}{{\algorithmicend\ }}
\algtext*{Indent}
\algtext*{EndIndent}

\usepackage{amsthm}
\newtheorem{theorem}{Theorem}

\theoremstyle{remark}

\newtheorem{remark}{Remark}

\DeclareMathOperator*{\argmax}{\arg\max}
\newcommand{\bbC}{\mathbb{C}}
\newcommand{\bbR}{\mathbb{R}}
\newcommand{\CN}{\mathcal{CN}}

\newcommand{\cK}{\mathcal{K}}

\newcommand{\cN}{\mathcal{N}}

\newcommand{\cS}{\mathcal{S}}

\newcommand{\tr}{\operatorname{tr}}

\begin{document}

\title{Energy-Efficient Tri-Hybrid Beamforming with\\Movable Pattern-Reconfigurable Antennas}

\author{Hong-Bae Jeon,~\IEEEmembership{Member,~IEEE}, Suhwan Jang, and Chan-Byoung Chae,~\IEEEmembership{Fellow,~IEEE}%
\thanks{\textit{(Corresponding Author: Suhwan Jang and Chan-Byoung Chae)}}%
\thanks{H.-B. Jeon is with the School of Electronic Engineering, Soongsil University, Seoul, Korea (e-mail: hongbae08@ssu.ac.kr).}%
\thanks{S. Jang is with the Department of Electrical and Computer Engineering, Ajou University, Suwon, Korea (e-mail: swanjang@ajou.ac.kr).}
\thanks{C.-B. Chae is with the School of Integrated Technology, Yonsei University, Seoul, Korea (e-mail: cbchae@yonsei.ac.kr).}
}

\maketitle
\begin{abstract}
Conventional hybrid precoding is confined to the digital/analog domains, leaving the spatial and radiation-pattern degrees-of-freedoms (DoFs) largely untapped. To unlock these, we propose a movable pattern-reconfigurable antenna (MPRA) and an energy-efficient tri-hybrid framework that jointly optimizes the precoders, antenna positions, and radiation-pattern states. We develop a hardware-inspired power model capturing the energy and transmission-time costs of antenna reconfiguration. The resulting mixed discrete-continuous fractional problem is solved via an alternating-optimization framework combining fractional programming, Riemannian optimization, and switching-aware pattern optimization. Simulations demonstrate near-exhaustive-search performance and consistent energy-efficiency (EE) gains, validating the complementary benefits of antenna movement, pattern reconfiguration, and tri-hybrid precoding under hardware-inspired power and latency costs.
\end{abstract}

\begin{IEEEkeywords}
Movable pattern-reconfigurable antenna (MPRA), tri-hybrid beamforming, energy-efficiency.
\end{IEEEkeywords}

\IEEEpeerreviewmaketitle
\section{Introduction}
\label{sec:intro}
\lettrine{T}{he} evolution of multiple-input multiple-output (MIMO) technologies for sixth-generation (6G) and beyond wireless network has continuously expanded the spatial degree-of-freedom (DoF) available for wireless transmission~\cite{elmimotut, sammag}. Nevertheless, conventional fixed-position antenna (FPA) arrays employ predetermined element locations that cannot adapt to the propagation environment, thereby limiting their ability to exploit spatial channel variations. To overcome this limitation, reconfigurable antenna technologies, such as movable antenna (MA)~\cite{mamag, rzhangma} and fluid antenna systems (FAS)~\cite{FAS, fas6gtuo}, have emerged as promising approaches for exploiting spatial channel variations. In representative position-reconfigurable FAS models, selecting favorable antenna locations, termed ports, provides spatial diversity~\cite{FAS, HBFASARIS}. Such spatial selection can also mitigate multi-user interference by exploiting locations where interfering signals undergo deep fading~\cite{vcfas22}. In MA systems, spatial reconfiguration is implemented by adaptively adjusting antenna positions within prescribed regions according to the propagation environment~\cite{mamag, Zhu2024Modeling}. This enables MA to reshape the effective channel, circumvent unfavorable fading regions, and consequently enhance MIMO performance~\cite{Ma2024MIMOCapacity}.

Recent studies have demonstrated substantial performance gains of MA over FPA. In~\cite{Zhu2024Modeling}, a field-response channel model characterized spatial multipath variations and revealed the fundamental spatial-diversity gains of MA. Building on this foundation,~\cite{Ma2024MIMOCapacity} characterized MA-MIMO capacity by jointly optimizing transmit/receive antenna positions and transmit covariance. For multi-user systems,~\cite{Zhu2024Multiuser} substantially reduced transmit power through joint antenna-position, power-control, and receive-combining optimization. In array processing,~\cite{Ma2024MultiBeam} jointly optimized antenna positions and weights to enhance desired multi-beam gains and suppress interference toward undesired directions. For efficient position optimization,~\cite{Liu2026DiscreteSampling} developed a discrete-sampling framework with sequential updates and Gibbs sampling to achieve near-optimal performance at low complexity, while~\cite{Zhang2026Evolutionary} used differential evolution and adaptive constraint penalties to reduce the complexity of conventional population-based methods. 

The synergy between MA and other technologies has also attracted growing attention. In~\cite{Wei2025MAIRS}, MA positioning was jointly designed with intelligent reflecting surface (IRS) beamforming, revealing that the additional gain offered by antenna mobility depends on the channel-reconfiguration capability of the IRS. Furthermore,~\cite{Ma2026ISAC} developed a large-timescale MA positioning based on statistical channel-state information (CSI), demonstrating a communication-sensing tradeoff through reduced inter-user channel correlation and sensing ambiguity. In the context of integrated-sensing-and-communications (ISAC),~\cite{Lyu2025ISAC} jointly optimized antenna positions and beamforming to enhance the ISAC performance. Overall, these studies have demonstrated the substantial benefits of MA, establishing it as a promising reconfigurable architecture for 6G and beyond.


Despite these advantages, conventional MA primarily exploit spatial DoF through antenna repositioning. With fixed antenna radiation characteristics, however, repositioning changes propagation distances and path-dependent spatial phases~\cite{Ma2024MIMOCapacity}. Position optimization alone therefore cannot directly reshape the radiation pattern to favor useful propagation paths or suppress interfering directions. This limitation is particularly relevant in multi-user and multipath environments, where paths arrive or depart along distinct angular directions~\cite{hjdrl, Zheng2025TriHybrid}.

To handle this, pattern-reconfigurable antenna (PRA) offers complementary electromagnetic (EM)-domain DoF by dynamically adjusting their radiation patterns~\cite{ramag}, redistributing directional gains to enhance desired paths and suppress less useful or interfering directions~\cite{tapbeamp, wenruiant}. Recent studies have demonstrated that PRA can substantially improve the performance of MIMO systems; 
In~\cite{Wang2025ERFAS}, joint analog-phase and radiation-state optimization on FAS improved spectral-efficiency (SE), received power, and bit-error-rate (BER), as validated by full-wave simulations and hardware experiments. In~\cite{Ying2025RmMIMO}, spherical-harmonic-based radiation-pattern synthesis was jointly configured with digital precoding, achieving significant SE gains with low channel-estimation overhead. These results highlight PRA as a promising means of adapting the angular response to the propagation environment.

Realizing the benefits of PRA requires jointly optimizing their radiation states with multiple precoding stages. In hardware-efficient multi-antenna systems, precoding is conventionally implemented through hybrid beamforming across the digital baseband and analog radio-frequency (RF) domains~\cite{hbyx}. Building upon this two-layer architecture, additional performance gains can be achieved by explicitly incorporating the antenna-domain DoFs provided by PRA, which has motivated the emergence of tri-hybrid beamforming~\cite{trimimo22}, which augments the hybrid layers with an EM-domain precoding layer~\cite{trimimo}. By configuring the antenna characteristics, it directly shapes the radiated field and further reduces reliance on costly RF-domain hardware~\cite{Zheng2025TriHybrid,trirad,ttbd}, thereby enabling flexible and energy-efficient transmission.

Furthermore, PRA naturally complements MA: MA selects favorable spatial locations, while PRA adapts the angular response at each location. Since both reshape the effective multi-user channel, their gains are best realized through joint design with precoders. Tri-hybrid beamforming therefore supports this integration by combining digital/analog RF processing with EM-domain antenna-state control, alongside spatial position adaptation. However, this couples antenna positions, radiation states, and hybrid precoders through the multipath channel, while movement and pattern switching incur time and energy costs. Existing studies largely address these DoFs separately, leaving their joint optimization under hardware-inspired reconfiguration costs unexplored. Although~\cite{Chen2025REMAA} electronically emulates movement through discrete radiation-position selection, it neither treats physical displacement and pattern shaping as independent DoFs nor accounts for multilayer precoding and realistic reconfiguration overhead.

Motivated by these considerations, we introduce a movable pattern-reconfigurable antenna (MPRA) architecture and develop an energy-efficient tri-hybrid framework for multi-user system that explicitly accounts for hardware-inspired reconfiguration delays and energy costs. We formulate and solve an energy-efficiency (EE) maximization problem that balances the channel gains offered by spatial and pattern reconfiguration against their hardware-inspired costs. The main contributions are summarized as follows:
\begin{itemize}
\item We propose an MPRA-enabled tri-hybrid transmission architecture that jointly leverages antenna positioning, radiation-pattern reconfiguration, and digital-analog precoding. A unified position- and pattern-dependent field-response model is established, in which antenna displacement controls the path-dependent spatial phases, whereas pattern selection reshapes the path-dependent directional gains. The resulting cascade explicitly incorporates pattern selection as an additional EM-domain precoding layer and establishes the tri-hybrid architecture.

\item We formulate a hardware-inspired EE model that captures both the energy expenditure and the loss of data-transmission time caused by antenna reconfiguration. We further demonstrate that operating each antenna at its maximum feasible synchronized speed is EE-optimal, thereby eliminating the movement-speed variables. Based on this model, the joint tri-hybrid MPRA design is formulated as a nonsmooth mixed discrete-continuous fractional program.

\item We develop an alternating optimization (AO) framework tailored to the problem. The hybrid-precoder block combines Dinkelbach's method with weighted minimum mean-squared error (WMMSE) optimization, using closed-form digital updates with bisection and Riemannian conjugate-gradient (RCG) analog updates. Thereafter, a switching-aware cyclic search optimizes discrete patterns while accounting for transition costs and avoiding an exponential number of state enumeration. The antenna-position optimization block uses a quadratic transform and WMMSE-based successive convex approximation (SCA) to yield convex subproblems. 

\item We establish that every block update preserves feasibility and does not decrease the original EE; consequently, the resulting EE sequence is monotonically nondecreasing and convergent. Numerical results further show that the proposed framework closely approaches the brute-force search (BFS) benchmark and consistently outperforms baselines in various scenarios.
\end{itemize}
%
%

\section{System Model and Problem Formulation}
\label{sec:system_model}

\subsection{Network Architecture}
\label{subsec:architecture}
\begin{figure}[t]
    \centering
    \includegraphics[width=0.99\linewidth]{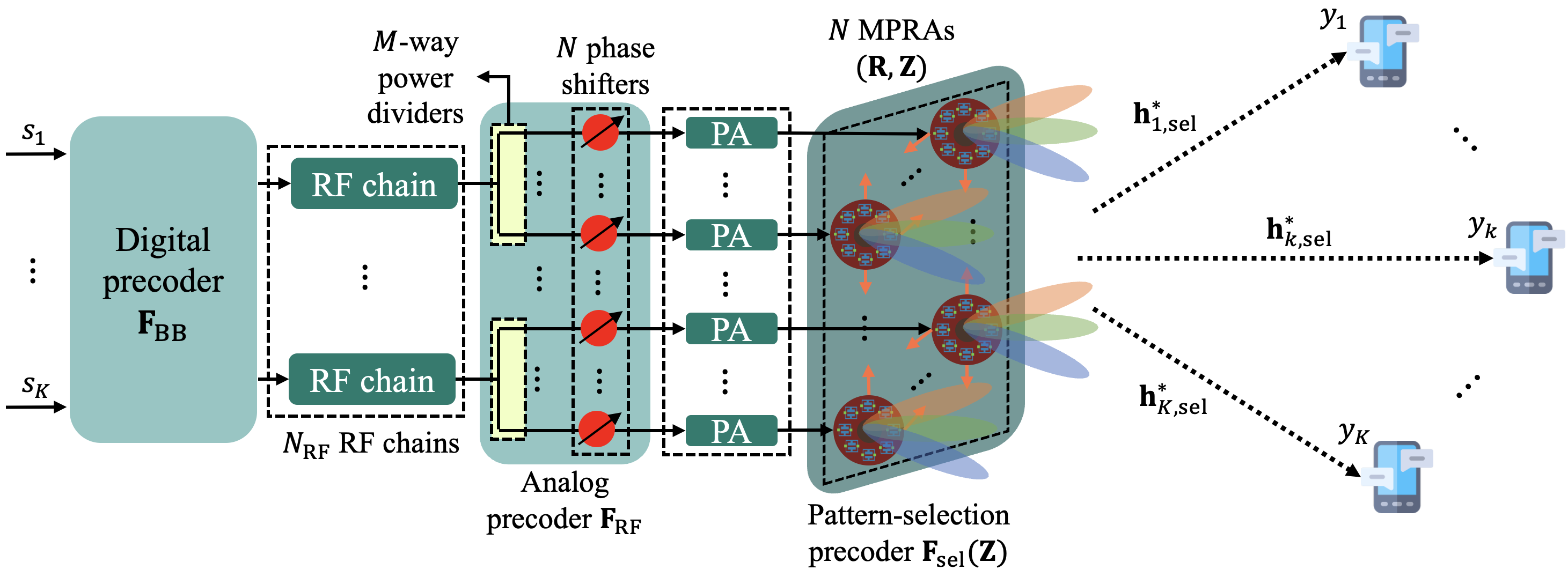}
    \caption{Illustration of a multi-user downlink system employing a tri-hybrid MPRA architecture at BS.}
    \label{fig_sys}
\end{figure}
As illustrated in Fig.~\ref{fig_sys}, we consider a multi-user downlink system in which a base station (BS) with $N_{\rm RF}$ RF chains and $N$ MPRAs simultaneously serves $K$ single-antenna users. 
The user, MPRA, RF chains, and radiation-pattern index sets are denoted by $\cK\triangleq\{1,\cdots,K\}$, $\cN\triangleq\{1,\cdots,N\}$, $\mathcal N_{\rm RF}=\{1, \cdots, N_{\rm RF}\}$, and $\cS\triangleq\{1,\cdots,S\}$, respectively, where $ K\leq N_{\rm RF}\leq N$. 
Let $\mathbf{s}=[s_1 \cdots s_K]^{\mathrm T}\sim\CN(\mathbf{0},\mathbf{I}_K)$ denote the data-symbol vector. The digital/analog precoders are denoted by $ \mathbf{F}_{\rm BB}=[\mathbf{f}_1 \cdots \mathbf{f}_K]\in\bbC^{N_{\rm RF}\times K}$ and $ \mathbf{F}_{\rm RF}\in\bbC^{N\times N_{\rm RF}}$, respectively. The transmit vector delivered to the $N$ MPRA ports is therefore $\mathbf{x}=\mathbf{F}_{\rm RF}\mathbf{F}_{\rm BB}\mathbf{s}$, with: 
\begin{equation}
\mathbb E[\|\mathbf x\|_2^2]
 =\|\mathbf{F}_{\rm RF}\mathbf{F}_{\rm BB}\|_{F}^{2}\le P_{\max},
 \label{eq:accepted_power}
\end{equation}
with power constraint $P_{\max}$. We assume a partially-connected phase-shifter network, where each MPRA is connected to exactly one RF chain~\cite{hbyx}. Let \(M\triangleq N/N_{\rm RF}(\in\mathbb Z)\) denote the number of MPRAs connected to each RF chain and define $\mathcal N_r
\triangleq
\{(r-1)M+1,\cdots,rM\}~(r\in\mathcal N_{\rm RF})$.\footnote{Herein for $N/N_{\rm RF}\notin \mathbb Z$, each RF chain is assigned $q=\lfloor N/N_{\rm RF}\rfloor$ MPRAs, with one additional MPRA assigned to each of $p=N-qN_{\rm RF}$ RF chains. The proposed framework can naturally be extended to these unequal-sized subarrays~\cite{Nguyen2020Unequal}; details are omitted for brevity and the tendency is shown in Section~\ref{subsec:performance_comparisons}.} The corresponding $\mathbf F_{\rm RF}$-feasible set is~\cite{hbyx}:
\begin{equation}
\begin{aligned}
 \mathcal F_{\rm RF}
 \triangleq
 \Bigg\{&
 \mathbf F_{\rm RF}=\mathrm{blkdiag}(\mathbf f_1, \cdots, \mathbf f_{N_{\rm RF}})\in\mathbb C^{N\times N_{\rm RF}},\mathbf f_r \in\mathbb C^{M},\\
& |[\mathbf f_r]_m|=\frac{1}{\sqrt M}~(m=1, \cdots, M)\bigg\}.
 \label{eq:analog_set}
\end{aligned}
\end{equation}
Since $ \mathbf F_{\rm RF}^{*}\mathbf F_{\rm RF}
 =\mathbf I_{N_{\rm RF}}$ by~\eqref{eq:analog_set},~\eqref{eq:accepted_power} becomes
\begin{equation}
\label{hcpc}
\|\mathbf{F}_{\rm RF}\mathbf{F}_{\rm BB}\|_{ F}^{2}=\|\mathbf{F}_{\rm BB}\|_{F}^{2}\le P_{\max}.
\end{equation}
\subsection{Movable Pattern-Reconfigurable Antenna (MPRA) Model}
\label{subsec:mpra_model}
The $n$th MPRA moves inside $\{\mathcal{C}_n\subset\bbR^2\}$, a non-overlapping $A\times A$ squares to prohibit physical collisions~\cite{Ma2024MIMOCapacity, Zhang2025MAHybrid}. Its given original and target positions are denoted by $\mathbf{r}_n^{0}=[x_n^0~y_n^0]^{\mathrm T}$ and $\mathbf{r}_n=[x_n~y_n]^{\mathrm T}$, respectively, and let $ \mathbf{R}^0\triangleq[\mathbf{r}_1^0 \cdots \mathbf{r}_N^0]\in\bbR^{2\times N}$ and $ \mathbf{R}\triangleq[\mathbf{r}_1 \cdots \mathbf{r}_N]\in\bbR^{2\times N}$. 
To prevent excessive EM coupling~\cite{Wei2026Mechanical, Ma2024MIMOCapacity}, $\{\mathbf r_n\}$ satisfy
\begin{equation}
 \mathbf{r}_n\in\mathcal{C}_n~ (\forall n\in\cN),
 ~
 \|\mathbf{r}_n-\mathbf{r}_m\|_2\geq d_{\min}~ (\forall n<m),
 \label{eq:position_constraints}
\end{equation}
with minimum separation $d_{\rm min}>0$ and so for $\{\mathbf r_n^0\}$ but $>d_{\min}$ instead of $\ge$; this ensures the feasible initialization developed in Section~\ref{subsec:overall_ao}. 

For pattern reconfiguration, each MPRA selects a pattern from a calibrated codebook of size $S$~\cite{Zheng2025TriHybrid, Wang2025ERFAS, tapbeamp, ttbd}:\footnote{Following the magnitude-gain normalization within unit sphere $\mathbb S^2$~\cite{Zheng2025TriHybrid}, each candidate satisfies $ \frac{1}{4\pi}\int_{\mathbb S^2}
 \overline{\psi}_s^2(\boldsymbol{\Omega})
  \mathrm d\boldsymbol{\Omega}
 =1~(\forall s\in\mathcal S)$.}
\begin{equation}
 \overline{\boldsymbol{\psi}}(\boldsymbol{\Omega})
 \triangleq
 [\overline{\psi}_1(\boldsymbol{\Omega}) \cdots
  \overline{\psi}_S(\boldsymbol{\Omega})]^{\mathrm T}, 
 \label{effp}
\end{equation}
where $\boldsymbol{\Omega}=(\theta,\phi)$ denotes an elevation and azimuth departure direction, respectively, and $\overline{\psi}_s(\boldsymbol{\Omega})\in\mathbb R_{+}$ denotes the $s$th normalized field-amplitude response,  
shared by all MPRAs.\footnote{We assume ideal port matching and neglect configuration-dependent mutual coupling~\cite{Zheng2025TriHybrid, Wang2025ERFAS, ttbd}, whereas the efficiency variations among the pattern candidates are neglected for ideal case.} 
With state selection $\mathbf z_n\in\{0,1\}^{S}$, the configured pattern is
\begin{equation}
 \psi_n(\boldsymbol{\Omega};\mathbf z_n)
 =
 \overline{\boldsymbol{\psi}}^{\mathrm T}(\boldsymbol{\Omega})
 \mathbf z_n,
 ~
 \mathbf 1_S^{\mathrm T}\mathbf z_n=1,
 \label{eq:pattern_selection}
\end{equation}
and we collect $\{\mathbf z_n\}$ in $\mathbf Z\triangleq[\mathbf z_1\cdots\mathbf z_N]
\in\{0,1\}^{S\times N}$.

\subsection{Tri-Hybrid Architecture with $\mathbf R$- and $\mathbf Z$-Dependency}
\label{subsec:channel}
We adopt a narrowband geometric block-fading model~\cite{Zheng2025TriHybrid, hjdrl, sjangbsq}. The channel between the BS and user $k$, varies with \((\mathbf R,\mathbf Z)\)~\cite{Ma2024MIMOCapacity,Zhang2025MAHybrid, Wei2026Mechanical}, consists of $L_k$ far-field propagation paths. The $\ell$th path is characterized by $\boldsymbol{\Omega}_{k,\ell}=(\theta_{k,\ell},\phi_{k,\ell})$, i.i.d. channel gain $\alpha_{k,\ell}\in\mathbb C$ with $\mathbb E[|\alpha_{k,\ell}|^2]=1$, and path-loss $\beta_k$~\cite{Zheng2025TriHybrid}. 
For $\boldsymbol{\Omega}=(\theta,\phi)$, define $ \mathbf u(\boldsymbol{\Omega})
 =
 [\sin\theta\cos\phi~
  \sin\theta\sin\phi~
  \cos\theta]^{\mathrm T}$. We use $k_0\triangleq2\pi/\lambda$ for the wavenumber, and embed $\mathbf r_n$ into $ \widetilde{\mathbf r}_n
 \triangleq
 [\mathbf r_n^{\mathrm T}~0]^{\mathrm T}$. 
 The antenna-wise pattern selections are collected into:
\begin{equation}
 \mathbf F_{\rm sel}(\mathbf Z)
 \triangleq
 \operatorname{blkdiag}
 \left\{
  \mathbf z_1,\cdots,\mathbf z_N
 \right\}
 \in\{0,1\}^{NS\times N}.
 \label{eq:pattern_selection_matrix}
\end{equation}
For \((k,\ell)\), define the $\mathbf R$-dependent spatial-response as $ \mathbf a_{k,\ell}(\mathbf R)
 \triangleq
 \begin{bmatrix}
 e^{-j k_0\mathbf u^{\mathrm T}
 (\boldsymbol{\Omega}_{k,\ell})
 \widetilde{\mathbf r}_1}\\
 \vdots\\
 e^{-j k_0\mathbf u^{\mathrm T}
 (\boldsymbol{\Omega}_{k,\ell})
 \widetilde{\mathbf r}_N}
 \end{bmatrix}
 \in\mathbb C^{N}$. To represent all \(S\) states of every MPRA before pattern selection, introduce 
$ \overline{\mathbf g}_{k,\ell}
 \triangleq
 \mathbf 1_N\otimes
 \overline{\boldsymbol{\psi}}
 (\boldsymbol{\Omega}_{k,\ell})
 \in\mathbb R_{+}^{NS}$ and $ \overline{\mathbf a}_{k,\ell}(\mathbf R)
 \triangleq
 \mathbf a_{k,\ell}(\mathbf R)
 \otimes\mathbf 1_S
 \in\mathbb C^{NS}$. Their Hadamard product $\overline{\mathbf v}_{k,\ell}(\mathbf R)
 \triangleq
 \overline{\mathbf g}_{k,\ell}
 \odot
 \overline{\mathbf a}_{k,\ell}(\mathbf R)
 \in\mathbb C^{NS}$ gives 
the pattern-expanded row channel: 
\begin{equation}
\mathbf h_{k,\rm sel}^{*}(\mathbf R)
 \triangleq
 \sqrt{\frac{\beta_k}{L_k}}
 \sum_{\ell=1}^{L_k}
 \alpha_{k,\ell}
 \overline{\mathbf v}_{k,\ell}^{\mathrm T}(\mathbf R)
 \in\mathbb C^{1\times NS}.
 \label{eq:expanded_path_response}
\end{equation}
The physical \(N\)-dimensional channel after pattern selection can then be factorized as $ \mathbf h_k^{*}(\mathbf R,\mathbf Z)
 =
 \mathbf h_{k,\rm sel}^{*}(\mathbf R) \mathbf F_{\rm sel}(\mathbf Z)
 \in\mathbb C^{1\times N}$. Equivalently, its \(n\)th entry is
\begin{equation}
 \big[\mathbf h_k^{*}(\mathbf R,\mathbf Z)\big]_n
 =
 \sqrt{\frac{\beta_k}{L_k}}
 \sum_{\ell=1}^{L_k}
 \alpha_{k,\ell}
 \overline{\boldsymbol{\psi}}^{\mathrm T}
 (\boldsymbol{\Omega}_{k,\ell})\mathbf z_n 
 e^{-j k_0\mathbf u^{\mathrm T}
 (\boldsymbol{\Omega}_{k,\ell})
 \widetilde{\mathbf r}_n}.
 \label{eq:mpra_channel}
\end{equation}
Finally, stacking the user channels gives $ \mathbf H_{\rm sel}(\mathbf R)
 \triangleq
 \begin{bmatrix}
  \mathbf h_{1,\rm sel}^{*}(\mathbf R)\\
  \vdots\\
  \mathbf h_{K,\rm sel}^{*}(\mathbf R)
 \end{bmatrix}
 \in\mathbb C^{K\times NS}$.\footnote{We assume that the geometric channel parameters $\{\beta_k,\alpha_{k,\ell},\boldsymbol{\Omega}_{k,\ell}\}$ are acquired at the beginning of each coherence block using CSI acquisition for MA~\cite{sjangma, sjangma22, jangcho}, and remain invariant during $T_c$-length coherence block~\cite{Ma2024MIMOCapacity,Zhang2025MAHybrid, Wei2026Mechanical}. The calibrated magnitude-pattern codebook is obtained offline and stored at BS~\cite{Wang2025ERFAS,Zheng2025TriHybrid}. Consequently, the BS can evaluate~\eqref{eq:tri_hybrid_signal_model} at candidate antenna positions and pattern states without estimating an independent $N$-dimensional channel for every candidate configuration.}
Consequently, the overall downlink channel exhibits the following tri-hybrid cascade:
\begin{equation}
 \mathbf y
 =
\underbrace{ \mathbf H_{\rm sel}(\mathbf R)
 \mathbf F_{\rm sel}(\mathbf Z)}_{\triangleq  \mathbf H_{\rm eff}(\mathbf R, \mathbf Z)}
 \mathbf F_{\rm RF}
 \mathbf F_{\rm BB}\mathbf s
 +
 \mathbf n.
 \label{eq:tri_hybrid_signal_model}
\end{equation}
Herein, with $(\mathbf F_{\rm BB}, \mathbf F_{\rm RF})$, $\mathbf F_{\rm sel}(\mathbf Z)$ offers the third precoding layer: switches the path amplitudes through the selected directional magnitude-gain patterns in~\eqref{effp}~\cite{trimimo, trimimo22}.\footnote{In~\eqref{eq:tri_hybrid_signal_model}, the equivalent tri-hybrid-domain signal $\overline {\mathbf x}= \mathbf F_{\rm sel}(\mathbf Z)\mathbf F_{\rm RF} \mathbf F_{\rm BB}\mathbf s$ satisfies
\(\mathbb E[\|\overline{\mathbf x}\|_2^2]=\|\mathbf F_{\rm sel}(\mathbf Z)\mathbf F_{\rm RF}
\mathbf F_{\rm BB}\|_{\rm F}^{2}
=\|\mathbf F_{\rm RF}\mathbf F_{\rm BB}\|_{\rm F}^{2} \le P_{\max}\) since \(\mathbf F_{\rm sel}^{*}(\mathbf Z)
\mathbf F_{\rm sel}(\mathbf Z)=\mathbf I_N\), same as~\eqref{eq:accepted_power}. Hence, we impose the power constraint at the physical MPRA ports, before the pattern-selection operator.} 

\subsection{Received Signal and Achievable SE}
\label{subsec:received_signal}
From~\eqref{eq:tri_hybrid_signal_model}, the received signal at user $k$ is
\begin{equation}
\begin{aligned}
 y_k
 =&\underbrace{\mathbf h_{k,\rm sel}^{*}(\mathbf R) \mathbf F_{\rm sel}(\mathbf Z)
 \mathbf{F}_{\rm RF}\mathbf{f}_k s_k}_{\text{desired signal}}\\
& +\underbrace{\sum_{j\neq k}\mathbf h_{k,\rm sel}^{*}(\mathbf R) \mathbf F_{\rm sel}(\mathbf Z)
 \mathbf{F}_{\rm RF}\mathbf{f}_j s_j}_{\text{multi-user interference}}+n_k,
 \label{eq:received_signal}
\end{aligned}
\end{equation}
where $n_k\sim\CN(0,\sigma_k^2)$. Hence, the SINR of user $k$ is
\begin{equation}
 \gamma_k
 =\frac{\left|\mathbf h_{k,\rm sel}^{*}(\mathbf R) \mathbf F_{\rm sel}(\mathbf Z)
 \mathbf{F}_{\rm RF}\mathbf{f}_k\right|^2}
 {\sum_{j\neq k}
 \left|\mathbf h_{k,\rm sel}^{*}(\mathbf R) \mathbf F_{\rm sel}(\mathbf Z)
 \mathbf{F}_{\rm RF}\mathbf{f}_j\right|^2+\sigma_k^2},
 \label{eq:sinr}
\end{equation}
and the sum-SE is therefore $ R_{\rm w} \triangleq\sum_{k=1}^{K}\log_2\left(1+\gamma_k\right)$.
 
\subsection{Reconfiguration Overhead and Block Energy}
\label{subsec:overhead_energy}
\subsubsection{Reconfiguration Overhead and Data-Transmission Time}
The physical movement of MPRA consumes a portion of each coherence block, giving rise to a tradeoff between the channel improvement obtained through reconfiguration and the remaining time available for data transmission~\cite{Wei2026Mechanical, Hu2026HowLong}. 
At the beginning, the BS occupies a known original configuration $ \mathbf R^0$ and $
 \mathbf Z^0
 \triangleq
 [\mathbf z_1^0\cdots\mathbf z_N^0]\in \{0, 1\}^{S\times N}$. 
 After channel acquisition and optimization, the MPRAs reconfigure from \(\mathbf R^0, \mathbf Z^0\) to \(\mathbf R, \mathbf Z\) before data transmission begins.

To explicitly account for the biaxial positioning, let \(r_{n,a}\) denote the \(a(\in\mathcal A)\)th coordinate of \(\mathbf r_n\). Extending~\cite{Wei2026Mechanical}, each MPRA is mounted on a Cartesian positioning stage driven by two orthogonal linear actuators. Herein, the two axes are synchronized to generate a constant-speed straight-line trajectory from \(\mathbf r_n^0\) to \(\mathbf r_n\)~\cite{Xiao2005Synchronization, motorbook}.\footnote{Note that the constant-speed approximation neglects the comparatively short acceleration and deceleration transients~\cite{Wei2026Mechanical}.} Define the axis-wise displacement and the travel distance respectively as $  d_{n,a}(\mathbf R)
 \triangleq
 \left|r_{n,a}-r_{n,a}^0\right|~(a\in\mathcal A)$ and $ D_n(\mathbf R)
 \triangleq
 \left\|\mathbf r_n-\mathbf r_n^0\right\|_2
 =
 \sqrt{\sum_{a\in\mathcal A}d_{n,a}^2(\mathbf R)}$. For \(D_n(\mathbf R)>0\), let \(v_n>0\) denote the resultant linear speed of MPRA \(n\) along its trajectory, where \(v_n=0(\leftrightarrow \forall v_{n,a}=0)\) is adopted by convention when \(D_n(\mathbf R)=0\). The corresponding speed of actuator \(a\) is
\begin{equation}
 v_{n,a}(\mathbf R,v_n)
 \triangleq
 v_n\frac{d_{n,a}(\mathbf R)}{D_n(\mathbf R)}~( a\in\mathcal A, D_n (\mathbf R)>0),
 \label{eq:synchronized_axis_speed}
\end{equation}
which ensures that both axes reach their target coordinates simultaneously~\cite{motorbook}. Let \(v_{n,a}^{\max}>0\) denote the maximum admissible speed of actuator \(a\)~\cite{Wei2026Mechanical}. 
From~\eqref{eq:synchronized_axis_speed}, the feasible $v_n$ is bounded by $ 0<v_n
 \leq
 \overline v_n(\mathbf R)
 \triangleq
 \min_{a\in\mathcal A: d_{n,a}(\mathbf R)>0}
 \frac{D_n(\mathbf R)}
 {d_{n,a}(\mathbf R)}
 v_{n,a}^{\max}$. 
The movement time of MPRA \(n\) is therefore $ T_n(\mathbf R,v_n)
 \triangleq
 \begin{cases}
 \frac{D_n(\mathbf R)}{v_n}
 &(D_n(\mathbf R)>0)\\
 0
 &(D_n(\mathbf R)=0)
 \end{cases}.$ Assuming that all MPRA positioning stages operate in parallel, the overall mechanical reconfiguration time is
\begin{equation}
 T_{\rm mov}(\mathbf R,\mathbf v)
 =
 \max_{n\in\mathcal N}
 T_n(\mathbf R,v_n),
 ~
 \mathbf v\triangleq[v_1 \cdots v_N]^{\mathrm T}.
 \label{eq:movement_time}
\end{equation}

Thereafter, the number of MPRAs whose pattern state changes is $ N_{\rm sw}(\mathbf{Z})
 \triangleq\frac{1}{2}\sum_{n=1}^{N}\|\mathbf{z}_n-\mathbf{z}_n^{0}\|_1$.
Assuming simultaneous electronic switching, the pattern-switching delay is $ T_{\rm pat}(\mathbf{Z})
 =\tau_{\rm sw}\mathbf{1}_{N_{\rm sw}(\mathbf{Z})>0}$ with hardware switching time $\tau_{\rm sw}$. The total overhead and data-transmission time are thus respectively~\cite{Wei2026Mechanical}: $T_{\rm oh}(\mathbf{R},\mathbf{Z}, \mathbf v)
 =T_{\rm mov}(\mathbf{R}, \mathbf v)+T_{\rm pat}(\mathbf{Z})$ and $T_{\rm d}(\mathbf{R},\mathbf{Z}, \mathbf v)
 =T_c-T_{\rm oh}(\mathbf{R},\mathbf{Z}, \mathbf v)$.\footnote{We conservatively assume that mechanical movement, and pattern switching are sequential.}

\subsubsection{Total Block Energy}
For each actuator, we adopt the speed-dependent stepper-motor model and allow its parameters to depend on both the antenna and the motion axis~\cite{Wei2026Mechanical, motorbook}. In particular, the power consumption is given by $ P_{{\rm M},n,a}(v)
 =
 \frac{v}{\ell_{0,n,a}}
 \mathcal M_{n,a}\left(\frac{v}{\ell_{0,n,a}}\right)$, where \(\ell_{0,n,a}\) is the linear-displacement factor of the lead-screw transmission, such that \(v=\ell_{0,n,a}\omega\) with angular speed $\omega$, and \(\mathcal M_{n,a}(\omega)\) is the available motor torque at \(\omega\): $ \mathcal M_{n,a}(\omega)
 =
 \frac{p_{n,a}\psi_{{\rm M},n,a}V_{n,a}}
 {\sqrt{R_{{\rm m},n,a}^{2}
       +\omega^{2}L_{{\rm m},n,a}^{2}}}
 -
 \frac{p_{n,a}\omega\psi_{{\rm M},n,a}^{2}R_{{\rm m},n,a}}
 {R_{{\rm m},n,a}^{2}
       +\omega^{2}L_{{\rm m},n,a}^{2}}$, where $p_{n,a}$ is the number of rotor teeth, $\psi_{{\rm M},n,a}$ is the peak magnet flux, and $V_{n,a}$, $R_{{\rm m},n,a}$, and $L_{{\rm m},n,a}$ are the motor voltage, phase resistance, and phase inductance, respectively.  Applying $ P_{{\rm M},n,a}(v)$ independently to the two orthogonal actuators, the total movement energy is\footnote{The accel/deceleration energies are neglected under the short-transient approximation in~\cite{Wei2026Mechanical}. Idle and position-holding powers are assumed negligible owing to de-energized actuators and self-locking transmissions~\cite{Wei2026Mechanical, motorbook}.}
\begin{equation}
 E_{\rm mov}(\mathbf R, \mathbf v)
 =
 \sum_{n=1}^{N}
 T_n(\mathbf R,v_n)
 \sum_{a\in\mathcal A}
 P_{{\rm M},n,a}\left(v_{n,a}(\mathbf R, v_n)\right).
 \label{eq:movement_energy}
\end{equation}
The pattern-switching energy is $ E_{\rm pat}(\mathbf{Z})
 =\varepsilon_{\rm sw}N_{\rm sw}(\mathbf{Z})$, where $\varepsilon_{\rm sw}$ is the energy required to switch one MPRA state.
\begin{remark}
\label{rem:continuous_position}
A stepper-motor-driven positioning admits only discrete locations determined by its axis-wise step sizes \(d_{s,a}\)~\cite{Wei2026Mechanical,motorbook}. Thus, the implementable position set of MPRA \(n\) is $\mathcal C_n^{\rm step}
=
\mathcal C_n\cap
\left\{
\mathbf r_n^0+
\begin{bmatrix}
d_{s,x}q_x\\ d_{s,y}q_y
\end{bmatrix}
:
[q_x~q_y]^{\rm T}\in\mathbb Z^2
\right\}$. For tractability, we optimize \(\mathbf R\) over the continuous \(\{\mathcal C_n\}\), providing a lower-bound on the implementable discrete-position performance. The resulting \(\mathbf R^\star\) can subsequently be mapped onto $\mathcal C_n^{\rm step}$ through the maximum-likelihood (ML) criterion.

At \(f_c=3.55\)~GHz, for instance, \(\lambda\simeq84.5\)~mm, and the step size \(d_s=1.2\)~mm from~\cite{Wei2026Mechanical} is only \(0.014\lambda\), yielding a maximum two-dimensional error of \(d_s/\sqrt{2}\simeq0.010\lambda\). Although such fine resolution suggests a minor quantization loss, independent rounding may violate the spacing constraint, necessitating the discrete-position selection~\cite{Wei2025MAIRS,Liu2026DiscreteSampling}. We therefore leave a comprehensive fully-discrete MPRA design for future work; nevertheless, the close agreement with BFS in Section~\ref{subsec:reliability_optimality} indicates that the fully-discrete framework is expected to incur only marginal performance degradation.
\end{remark}

The data-transmit power is modeled as~\cite{trimimo, madc, trirad}:
\begin{equation}
 P_{\rm data}
 =\underbrace{\frac{\|\mathbf{F}_{\rm BB}\|_F^2}{\eta_{\rm PA}}}_{\triangleq P_{\rm PA}}+\underbrace{ P_{\rm LO}+N_{\rm RF}(2P_{\rm DAC} +P_{\rm RF}) +NP_{\rm PS}}_{\triangleq P_{\rm circ}},
 \label{eq:data_power}
\end{equation}
where $\eta_{\rm PA}\in(0,1]$ denotes the power-amplifier (PA) efficiency, and $P_{\rm LO}$, $P_{\rm PA}$, $P_{\rm DAC}$, $P_{\rm RF}$, and $P_{\rm PS}$ denote the local-oscillator (LO), PA, per digital-to-analog converter (DAC), per-RF-chain, and per-phase-shifter circuit powers, respectively~\cite[Table~II]{trimimo}. 
Hence, the total block energy is~\cite{Wei2026Mechanical}:
\begin{equation}
 E_{\rm blk}
 =E_{\rm mov}(\mathbf{R}, \mathbf v)+E_{\rm pat}(\mathbf{Z})+T_{\rm d}(\mathbf{R},\mathbf{Z}, \mathbf v)P_{\rm data}.
 \label{eq:block_energy}
\end{equation}

\subsection{Problem Formulation}
\label{sec:problem_formulation}
\label{subsec:ee_problem}
For a system bandwidth $W$, the EE is
\begin{equation}
 \eta_{\rm EE}
 =\frac{WT_{\rm d}(\mathbf{R},\mathbf{Z},\mathbf v) R_{\rm w}}
 {E_{\rm blk}}~{\rm [bit/J]}.
 \label{eq:energy_efficiency}
\end{equation}
We aim to maximize $ \eta_{\rm EE}$ within each coherence block by jointly designing $ \mathcal X\triangleq\{\mathbf{F}_{\rm BB},\mathbf{F}_{\rm RF},\mathbf{R},\mathbf{Z}, \mathbf v\}$:
\begin{equation}
\begin{aligned}
&\max_{\mathcal X\triangleq\{\mathbf{F}_{\rm BB},\mathbf{F}_{\rm RF},\mathbf{R},\mathbf{Z}, \mathbf v\}} 
 ~\eta_{\rm EE}(\mathcal X)
\\
\mathrm{s.t.}~&\eqref{hcpc},~\mathbf F_{\rm RF}
 \in\mathcal F_{\rm RF},~\mathbf z_n\in\{0,1\}^{S},~\mathbf 1_S^{\mathrm T}\mathbf z_n=1,~\eqref{eq:position_constraints}\\
 &v_n\in\mathcal V_n(\mathbf R)~(\forall n\in\mathcal N),~
 T_{\rm oh}(\mathbf R,\mathbf Z, \mathbf v)
 \leq T_c-\epsilon_T,\\
 \label{prob:ee_qos}
\end{aligned}
\end{equation}
where \(\epsilon_T>0\) guarantees a strictly positive $T_{\rm d}$ and $\mathcal V_n(\mathbf R)
\triangleq
\begin{cases}
\{0\},
&(D_n(\mathbf R)=0)\\
(0,\overline v_n(\mathbf R)]
&(D_n(\mathbf R)>0).
\end{cases}$ Problem~\eqref{prob:ee_qos} is a mixed-integer fractional programming, which is nonconvex due to the fractional objective involving both continuous and discrete variables, as well as their nonsmooth coupling through the reconfiguration-time and mechanical-energy models. Therefore, we develop an AO framework for solving~\eqref{prob:ee_qos}.

\section{Proposed EE Optimization Framework}
\label{sec:proposed_algorithm}

\subsection{Optimal $\{v_{n,a}\}$ of MPRA}
\label{subsec:optimal_speed}
We first examine $\{v_{n,a}\}$ for fixed \((\mathbf F_{\rm BB},\mathbf F_{\rm RF},\mathbf R,\mathbf Z)\). 
It is known that from~\cite{Wei2026Mechanical}, \(\mathcal M_{n,a}(\omega)\) is nonnegative and non-increasing over $0\leq\omega\leq\omega_{n,a}^{\max}$, where $\omega_{n,a}^{\max}=\frac{v_{n,a}^{\max}}{\ell_{0,n,a}}$. Thereafter, we can consider the following theorem:
\begin{theorem}
\label{prop:optimal_speed}
For every fixed \((\mathbf F_{\rm BB},\mathbf F_{\rm RF},\mathbf R,\mathbf Z)\), $\eta_{\rm EE}$ is non-decreasing with respect to each \(v_n\) over \(\mathcal V_n(\mathbf R)\):
\begin{equation}
 v_n^{\star}(\mathbf R)
 =
 \begin{cases}
 \overline v_n(\mathbf R)
 &(D_n(\mathbf R)>0)\\
 0
 &(D_n(\mathbf R)=0)
 \end{cases}~( n\in\mathcal N).
 \label{eq:optimal_speed}
\end{equation}
\end{theorem}
\begin{proof}
Consider an MPRA \(n\) satisfying \(D_n(\mathbf R)>0\). 
The contribution $ E_{{\rm mov},n}(\mathbf R,v_n)$ of MPRA \(n\) in~\eqref{eq:movement_energy} is
\begin{equation}
\begin{aligned}
 E_{{\rm mov},n}(\mathbf R,v_n)
     &\triangleq
     T_n(\mathbf R,v_n)
     \sum_{a\in\mathcal A}
     P_{{\rm M},n,a}\big(v_{n,a}(\mathbf R,v_n)\big)
     \\
 &=
 \sum_{\substack{a\in\mathcal A:\\d_{n,a}(\mathbf R)>0}}
 \frac{d_{n,a}(\mathbf R)}{\ell_{0,n,a}}
 \mathcal M_{n,a}
 \left(
 \frac{v_nd_{n,a}(\mathbf R)}
 {D_n(\mathbf R)\ell_{0,n,a}}
 \right).
 \label{eq:movement_energy_speed_form}
\end{aligned}
\end{equation}
Its derivative with respect to \(v_n\) is
\begin{equation}
 \frac{\partial E_{{\rm mov},n}}{\partial v_n}
 =
 \sum_{\substack{a\in\mathcal A:\\d_{n,a}(\mathbf R)>0}}
 \frac{d_{n,a}^{2}(\mathbf R)}
 {D_n(\mathbf R)\ell_{0,n,a}^{2}}
 \mathcal M_{n,a}'
 \left(
 \frac{v_nd_{n,a}(\mathbf R)}
 {D_n(\mathbf R)\ell_{0,n,a}}
 \right)
 \leq0,
 \label{eq:movement_energy_derivative}
\end{equation}
since $\mathcal M_{n,a}$ is non-increasing. 

Moreover, \(T_n(\mathbf R,v_n)=\frac{D_n(\mathbf R)}{v_n}\) is decreasing in \(v_n\). Thus, increasing \(v_n\) cannot grow $ T_{\rm mov}(\mathbf R,\mathbf v) =\max_{m\in\mathcal N}T_m(\mathbf R,v_m)$, and thus cannot decrease \(T_{\rm d}\). For fixed variables, \(R_{\rm w}\), \(P_{\rm data}\), \(T_{\rm pat}\), and \(E_{\rm pat}\) are independent of \(\mathbf v\). Dividing the numerator and denominator of~\eqref{eq:energy_efficiency} by \(T_{\rm d}>0\):
\begin{equation}
 \eta_{\rm EE}
 =
 \frac{W R_{\rm w}}
 {P_{\rm data}
 +\dfrac{E_{\rm mov}+E_{\rm pat}}{T_{\rm d}}}.
 \label{eq:normalized_ee}
\end{equation}
Since increasing \(v_n\) cannot increase \(E_{\rm mov}+E_{\rm pat}\) and cannot decrease \(T_{\rm d}\), it cannot increase the denominator. Hence, \(\eta_{\rm EE}\) is non-decreasing in \(v_n\), and its maximum over \(\mathcal V_n(\mathbf R)\) is attained at~\eqref{eq:optimal_speed}. For \(D_n(\mathbf R)=0\), it follows from \(\mathcal V_n(\mathbf R)=\{0\}\).
\end{proof}
Theorem~\ref{prop:optimal_speed} extends the result in~\cite{Wei2026Mechanical} to independently controlled MPRAs undergoing synchronized biaxial motion. By defining $ \widehat{\mathbf v}(\mathbf R)
 \triangleq
 [v_1^{\star}(\mathbf R) \cdots
  v_N^{\star}(\mathbf R)]^{\mathrm T}$ hereafter, $\mathbf v$ is set to
\(\mathbf v=\widehat{\mathbf v}(\mathbf R)\), and we use $ \widehat T_{\rm d}(\mathbf R,\mathbf Z)
 \triangleq
 T_{\rm d}\big(\mathbf R,\mathbf Z,
 \widehat{\mathbf v}(\mathbf R)\big)$ and $ \widehat E_{\rm rec}(\mathbf R,\mathbf Z)
 \triangleq
 E_{\rm mov}\big(\mathbf R,\widehat{\mathbf v}(\mathbf R)\big)
 +E_{\rm pat}(\mathbf Z)$ to denote the resulting $T_{\rm d}$ and total reconfiguration energy, respectively. Accordingly, $\eta_{\rm EE}$ becomes $ \eta_{\rm EE}
 =
 \frac{
 W\widehat T_{\rm d}(\mathbf R,\mathbf Z)
 R_{\rm w}
 }{
 \widehat E_{\rm rec}(\mathbf R,\mathbf Z)
 +\widehat T_{\rm d}(\mathbf R,\mathbf Z)P_{\rm data}
 }$.

\subsection{Dinkelbach-WMMSE Optimization of $(\mathbf F_{\rm BB}, \mathbf F_{\rm RF})$}
\label{subsec:dinkelbach_wmmse}
We next optimize
\((\mathbf F_{\rm BB},\mathbf F_{\rm RF})\) for fixed
\((\mathbf R,\mathbf Z)\). For brevity, define $ \mathsf N(\mathbf F_{\rm BB},\mathbf F_{\rm RF})
 \triangleq
 W\widehat T_{\rm d} R_{\rm w}$ and $ \mathsf D(\mathbf F_{\rm BB},\mathbf F_{\rm RF})
 \triangleq
 \widehat E_{\rm rec}
 +\widehat T_{\rm d}P_{\rm data}$, where the dependence of \(\widehat T_{\rm d}\) and \(\widehat E_{\rm rec}\) on \((\mathbf R,\mathbf Z)\) is temporarily suppressed since \((\mathbf R,\mathbf Z)\) are fixed. 
For a Dinkelbach parameter \(\lambda_{\rm D}\geq0\), consider $ \mathcal J_{\lambda_{\rm D}} \triangleq
 \mathsf N -\lambda_{\rm D}\mathsf D$. The classical Dinkelbach characterization states that the globally optimal EE \(\eta_{\rm EE, D}^{\star}\) satisfies $ \max_{\mathbf F_{\rm BB},\mathbf F_{\rm RF}}
 \left\{
 \mathsf N-\eta_{\rm EE, D}^{\star}\mathsf D
 \right\}
 =0$ over feasible \((\mathbf F_{\rm BB},\mathbf F_{\rm RF})\)~\cite{Din}. Since \(\widehat T_{\rm d}>0\) is constant herein, maximizing $\mathcal J_{\lambda_{\rm D}}$ is equivalent to maximizing
\begin{equation}
 W R_{\rm w}
 -\lambda_{\rm D}P_{\rm data},
 \label{eq:reduced_dinkelbach_precoder}
\end{equation}
since \(-\lambda_{\rm D}\widehat E_{\rm rec}/\widehat T_{\rm d}\)
is independent of \((\mathbf F_{\rm BB},\mathbf F_{\rm RF})\).

To handle the interference-coupled $R_w$, define 
$ \mathbf g_k^{*}
 \triangleq
 \mathbf h_{k,\rm sel}^{*}(\mathbf R)
 \mathbf F_{\rm sel}(\mathbf Z)
 \mathbf F_{\rm RF}
 \in\mathbb C^{1\times N_{\rm RF}}$. For equalizer \(u_k\in\mathbb C\), the estimate of \(s_k\) is
\(\widehat s_k=u_k^{*}y_k\), and the corresponding MSE is
\begin{equation}
\begin{aligned}
 e_k
 &\triangleq
 \mathbb E\left[
 \left|u_k^{*}y_k-s_k\right|^{2}
 \right]\\
 &=
 |u_k|^{2}
 \left(
 \sum_{j=1}^{K}
 \left|\mathbf g_k^{*}\mathbf f_j\right|^{2}
 +\sigma_k^{2}
 \right)
 -2\Re\left\{
 u_k^{*}\mathbf g_k^{*}\mathbf f_k
 \right\}
 +1.
 \label{eq:user_mse}
\end{aligned}
\end{equation}
Introducing an auxiliary MSE weight \(\varpi_k>0\), the WMMSE identity for SE gives~\cite{Zheng2025TriHybrid}:
\begin{equation}
 R_{\rm w}
 =
 \frac{1}{\ln2}
 \max_{\substack{\{u_k\}\\\{\varpi_k>0\}}}
 \sum_{k=1}^{K}
 \left(
 \ln\varpi_k-\varpi_ke_k+1
 \right).
 \label{eq:wmmse_identity}
\end{equation}
Hence, substituting~\eqref{eq:wmmse_identity} into~\eqref{eq:reduced_dinkelbach_precoder}: 
\begin{equation}
\begin{aligned}
&\max_{\substack{
 \mathbf F_{\rm BB},\mathbf F_{\rm RF},\\
 \{u_k\in\mathbb C, \varpi_k>0\}_{k=1}^{K}
 }}
 \frac{W}{\ln 2}
 \sum_{k=1}^{K}
 \left(
 \ln\varpi_k-\varpi_k e_k+1
 \right)
 -
 \frac{\lambda_{\rm D}}{\eta_{\rm PA}}
 \left\|\mathbf F_{\rm BB}
 \right\|_{ F}^{2}
 \\
& \mathrm{s.t.}
 \left\|\mathbf F_{\rm BB}
 \right\|_{ F}^{2}
 \leq P_{\max},~\mathbf F_{\rm RF}
 \in\mathcal F_{\rm RF}.
\end{aligned}
\label{prob:dinkelbach_wmmse}
\end{equation}

\subsubsection{Update of $\{u_k, \varpi_k\}$}
For fixed \((\mathbf F_{\rm BB},\mathbf F_{\rm RF})\), the optimal auxiliary variables of~\eqref{prob:dinkelbach_wmmse} are given by
\begin{equation}
 u_k^{\star}
 =
 \frac{\mathbf g_k^{*}\mathbf f_k}
 {\sum_{j=1}^{K}
 |\mathbf g_k^{*}\mathbf f_j|^{2}
 +\sigma_k^{2}},~\varpi_k^{\star}
 =
 \left(e_k^{\rm MMSE}\right)^{-1}
 =1+\gamma_k.
 \label{eq:mmse_weight}
\end{equation}

\subsubsection{Update of $\mathbf F_{\rm BB}$}
For fixed
\((\mathbf F_{\rm RF},\{u_k,\varpi_k\})\), define $\mathbf A
 \triangleq
 \sum_{k=1}^{K}
 \varpi_k|u_k|^{2}
 \mathbf g_k\mathbf g_k^{*}
 \succeq\mathbf0$, $\mathbf B
 \triangleq
 \left[
 \varpi_1u_1\mathbf g_1 
 \cdots 
 \varpi_Ku_K\mathbf g_K
 \right]$, and $\rho_{\rm D}
 \triangleq
 \frac{\lambda_{\rm D}\ln2}
 {W\eta_{\rm PA}}$. The $\mathbf F_{\rm BB}$-subproblem then becomes
\begin{equation}
\begin{aligned}
& \min_{\mathbf F_{\rm BB}}~
 \tr\left[
 \mathbf F_{\rm BB}^{*}
 \left(\mathbf A+\rho_{\rm D}\mathbf I\right)
 \mathbf F_{\rm BB}
 \right]
 -2\Re\left\{
 \tr(\mathbf B^{*}\mathbf F_{\rm BB})
 \right\}\\
& \mathrm{s.t.}~ \|\mathbf F_{\rm BB}\|_{ F}^{2}\leq P_{\max}.
 \label{prob:bb_update}
\end{aligned}
\end{equation}
Problem~\eqref{prob:bb_update} is a convex quadratic program with its solution:
\begin{equation}
 \mathbf F_{\rm BB}^{\star}(\mu)
 =
 \left[
 \mathbf A+
 \left(\rho_{\rm D}+\mu\right)\mathbf I
 \right]^{\dagger}
 \mathbf B,
 \label{eq:bb_closed_form}
\end{equation}
where \(\mu\geq0\) is chosen to satisfy $ \mu
 \left(
 \|\mathbf F_{\rm BB}^{\star}(\mu)\|_{ F}^{2}
 -P_{\max}
 \right)=0$. If the unconstrained solution at \(\mu=0\) violates the power budget, the unique \(\mu>0\) satisfying
\(\|\mathbf F_{\rm BB}^{\star}(\mu)\|_{ F}^{2}=P_{\max}\)
can be found by scalar bisection.

\subsubsection{Update of $\mathbf F_{\rm RF}$}
\label{subsubsec:riemannian_rf}
We next optimize $\mathbf F_{\rm RF}$ for fixed
\((\mathbf F_{\rm BB},\{u_k,\varpi_k\})\). Let \(r(n)\) denote
the unique RF chain connected to MPRA \(n\), i.e.,
\(n\in\mathcal N_{r(n)}\), and collect the \(N\) active analog
coefficients into $ \mathbf q
 \triangleq
 \sqrt M
 [[\mathbf F_{\rm RF}]_{1,r(1)}
 \cdots
 [\mathbf F_{\rm RF}]_{N,r(N)}]^{\mathrm T}$. The fixed zero entries of \(\mathbf F_{\rm RF}\) are excluded from
\(\mathbf q\). By~\eqref{eq:analog_set}, the feasible set of
\(\mathbf q\) is the product complex-circle manifold $ \mathcal M_{\rm cc}^{N}
 \triangleq
 \left\{
 \mathbf q\in\mathbb C^{N}:
 |q_n|=1~( \forall n\in\mathcal N)
 \right\}$. The mapping from \(\mathbf q\) to $\mathbf F_{\rm RF}$ is denoted by \(\mathbf F_{\rm RF}(\mathbf q)\), whose entries satisfy $ [\mathbf F_{\rm RF}(\mathbf q)]_{n,r}
 =
 \begin{cases}
 \frac{q_n}{\sqrt M}&(r=r(n))\\
 0&(r\neq r(n))
 \end{cases}.$
 
For notational convenience, 
define $ [\mathbf c_{k,j}]_n
 \triangleq
 \frac{
 [\mathbf h_k(\mathbf R,\mathbf Z)]_n
 [\mathbf f_j]_{r(n)}^{*}
 }{\sqrt M}~ (n\in\mathcal N)$. It then follows that $ \mathbf h_k^{*}(\mathbf R,\mathbf Z)
 \mathbf F_{\rm RF}(\mathbf q)\mathbf f_j
 =
 \mathbf c_{k,j}^{*}\mathbf q$. Thus, the terms of the Dinkelbach-WMMSE problem that depend on \(\mathbf q\) can then be:
\begin{equation}
 \min_{\mathbf q\in\mathcal M_{\rm cc}^{N}}
 f_{\rm RF}(\mathbf q)
 \triangleq
 \mathbf q^{*}\mathbf C\mathbf q
 -2\Re\{\mathbf d^{*}\mathbf q\},
 \label{prob:riemannian_rf}
\end{equation}
where $ \mathbf C
 \triangleq
 \sum_{k=1}^{K}
 \varpi_k|u_k|^{2}
 \sum_{j=1}^{K}
 \mathbf c_{k,j}\mathbf c_{k,j}^{*}
 \succeq\mathbf0$ and $ \mathbf d
 \triangleq
 \sum_{k=1}^{K}
 \varpi_ku_k\mathbf c_{k,k}$.
Problem~\eqref{prob:riemannian_rf} can therefore be addressed by 
an RCG method with Armijo backtracking~\cite{hbyx, arpo}. It generates a non-increasing and converging sequence of \(\{\mathbf q\}\) and terminates when $ \left\|
 \operatorname{grad}f_{\rm RF}(\mathbf q)
 \right\|_2
 \leq\epsilon_{\rm RF}$ with Riemannian gradient $ \operatorname{grad}$~\cite{hbyx}. The resulting $\mathbf F_{\rm RF}$ is then 
 \(\mathbf F_{\rm RF}=\mathbf F_{\rm RF}(\mathbf q)\).

\subsubsection{Dinkelbach-Parameter Update and Termination}
Let $i\geq0$ denote the Dinkelbach iteration index and initialize $\lambda_{\rm D}^{(0)}=
\frac{\mathsf N(\mathbf F_{\rm BB}^{(0)}, \mathbf F_{\rm RF}^{(0)})}
{\mathsf D(\mathbf F_{\rm BB}^{(0)}, \mathbf F_{\rm RF}^{(0)})}$. 
For a fixed $\lambda_{\rm D}^{(i)}$, we alternately update
$\{u_k^{(i)},\varpi_k^{(i)}\}_{k=1}^{K}$, $\mathbf F_{\rm BB}^{(i)}$, and
$\mathbf F_{\rm RF}^{(i)}$ according to~\eqref{eq:mmse_weight},
\eqref{eq:bb_closed_form}, and the RCG procedure in
Section~\ref{subsubsec:riemannian_rf}, respectively. Let
$\big(\mathbf F_{\rm BB}^{(i+1)},\mathbf F_{\rm RF}^{(i+1)}\big)$
denote the resulting feasible hybrid precoders. The associated Dinkelbach residual is $ \mathcal J_{\lambda_{\rm D}^{(i)}}
\left(
\mathbf F_{\rm BB}^{(i+1)},
\mathbf F_{\rm RF}^{(i+1)}
\right)=
 \mathsf N\left(
 \mathbf F_{\rm BB}^{(i+1)},
 \mathbf F_{\rm RF}^{(i+1)}
 \right)-\lambda_{\rm D}^{(i)}
 \mathsf D\left(
 \mathbf F_{\rm BB}^{(i+1)},
 \mathbf F_{\rm RF}^{(i+1)}
 \right)$. 
Thereafter, $ \lambda_{\rm D}^{(i)}$ is updated: $ \lambda_{\rm D}^{(i+1)}
 =
 \frac{
 \mathsf N\left(
 \mathbf F_{\rm BB}^{(i+1)},
 \mathbf F_{\rm RF}^{(i+1)}
 \right)}{
 \mathsf D\left(
 \mathbf F_{\rm BB}^{(i+1)},
 \mathbf F_{\rm RF}^{(i+1)}
 \right)}$. The overall Dinkelbach routine terminates when $
|\lambda_{\rm D}^{(i+1)}
-\lambda_{\rm D}^{(i)}|
\leq
\epsilon_{\rm D}$ for given tolerance $\epsilon_{\rm D}>0$. Since $\{u_k^{(i)},\varpi_k^{(i)}\}_{k=1}^{K}$ and $\mathbf F_{\rm BB}$-updates optimize their respective blocks and the Armijo RCG update does not decrease the objective~\cite{hbyx}, the resulting $(\mathbf F_{\rm BB}^{(i+1)}, \mathbf F_{\rm RF}^{(i+1)})$ satisfies
\begin{equation}
\mathcal J_{\lambda_{\rm D}^{(i)}}
\left(
\mathbf F_{\rm BB}^{(i+1)},
\mathbf F_{\rm RF}^{(i+1)}
\right)
\geq
\mathcal J_{\lambda_{\rm D}^{(i)}}
\left(
\mathbf F_{\rm BB}^{(i)},
\mathbf F_{\rm RF}^{(i)}
\right)
=0,
\label{eq:dinkelbach_inner_ascent}
\end{equation}
where the right-side equality follows from $\lambda_{\rm D}^{(i)}
=
\frac{
\mathsf N(\mathbf F_{\rm BB}^{(i)},\mathbf F_{\rm RF}^{(i)})
}{
\mathsf D(\mathbf F_{\rm BB}^{(i)},\mathbf F_{\rm RF}^{(i)})
}$:
\begin{equation}
\lambda_{\rm D}^{(i+1)}
-\lambda_{\rm D}^{(i)}
=
\frac{\mathcal J_{\lambda_{\rm D}^{(i)}}
\left(
\mathbf F_{\rm BB}^{(i+1)},
\mathbf F_{\rm RF}^{(i+1)}
\right)
}
{
\mathsf D(
\mathbf F_{\rm BB}^{(i+1)},
\mathbf F_{\rm RF}^{(i+1)})
}
\geq0,
\label{eq:dinkelbach_parameter_monotonicity}
\end{equation}
and EE generated by $(\mathbf F_{\rm BB}, \mathbf F_{\rm RF})$-block is nondecreasing.


\subsection{Switching-Aware Iterative Optimization of $\mathbf Z$}
\label{subsec:iterative_pattern_update}
We then optimize $\mathbf Z$ for fixed $\mathbf R$ and $(\mathbf F_{\rm BB}, \mathbf F_{\rm RF})$. Let \(\overline{\mathbf R}\) denote the position configuration held fixed during this $\mathbf Z$-update block. Unlike exhaustive search over all \(S^N\) configurations, we successively update one MPRA pattern at a time using a cyclic candidate-testing procedure inspired by~\cite{trirad}. 

\subsubsection{Candidate Pattern Configuration and Channel Update}
Let $ \mathbf Z^{[p]}= [\mathbf z_1^{[p]} \cdots \mathbf z_N^{[p]}]$ denote the pattern configuration after \(p\)th pattern-coordinate iteration, 
and let \(s_n^{[p]}\) be the unique index satisfying $ \mathbf z_n^{[p]}=\mathbf e_{s_n^{[p]}}$. For fixed \(\mathbf r_n\), define the channel column associated with MPRA \(n\) operating in \(s\): $ \mathbf h_{n,s}(\mathbf r_n)
 \triangleq
 \begin{bmatrix}
 \chi_{1,n,s}(\mathbf r_n)\\
 \vdots\\
 \chi_{K,n,s}(\mathbf r_n)
 \end{bmatrix}
 \in\mathbb C^K$, where $ \chi_{k,n,s}(\mathbf r_n)
 =
 \sqrt{\frac{\beta_k}{L_k}}
 \sum_{\ell=1}^{L_k}
 \alpha_{k,\ell}
 \overline\psi_s(\boldsymbol\Omega_{k,\ell})
 e^{-j
 \boldsymbol\kappa_{k,\ell}^{\mathrm T}\mathbf r_n}$ with $\boldsymbol\kappa_{k,\ell}
\triangleq
k_0
\begin{bmatrix}
[\mathbf u(\boldsymbol\Omega_{k,\ell})]_x\\
[\mathbf u(\boldsymbol\Omega_{k,\ell})]_y
\end{bmatrix}
\in\mathbb R^2$ is the channel contribution of MPRA \(n\) toward user \(k\). For \(\mathbf Z=[\mathbf e_{s_1} \cdots \mathbf e_{s_N}]\), $\mathbf H_{\rm eff}$ can be expressed:
\begin{equation}
 \mathbf H_{\rm eff}(\overline{\mathbf R},\mathbf Z)
 =
 \left[
 \mathbf h_{1,s_1}(\overline{\mathbf r}_1) 
 \cdots 
 \mathbf h_{N,s_N}(\overline{\mathbf r}_N)
 \right]
 \in\mathbb C^{K\times N}.
 \label{eq:effective_channel_column_form}
\end{equation}

At \(p\geq1\), the MPRA to be updated is selected cyclically as $ n_p
 \triangleq
 1+\operatorname{mod}(p-1,N)$. 
 For \(s\in\mathcal S\triangleq\{1,\cdots,S\}\), define
\begin{equation}
 \mathbf Z_{n_p\leftarrow s}^{[p-1]}
 \triangleq
 \big[
 \mathbf z_1^{[p-1]} \cdots 
 \mathbf z_{n_p-1}^{[p-1]}~
 \mathbf e_s~\mathbf z_{n_p+1}^{[p-1]} \cdots 
 \mathbf z_N^{[p-1]}
 \big].
 \label{eq:pattern_candidate_replacement}
\end{equation}
Since only the \(n_p\)th pattern state is replaced, substituting~\eqref{eq:pattern_candidate_replacement} into~\eqref{eq:effective_channel_column_form} gives
\begin{equation}
\begin{aligned}
 &\mathbf H_{\rm eff}
 \left(
 \overline{\mathbf R},
 \mathbf Z_{n_p\leftarrow s}^{[p-1]}
 \right)
 \\
 &=
 \mathbf H_{\rm eff}
 \left(
 \overline{\mathbf R},
 \mathbf Z^{[p-1]}
 \right)
 +
 \left[
 \mathbf h_{n_p,s}(\overline{\mathbf r}_{n_p})
 -
 \mathbf h_{n_p,s_{n_p}^{[p-1]}}
 (\overline{\mathbf r}_{n_p})
 \right]
 \mathbf e_{n_p}^{\mathrm T},
 \label{eq:candidate_channel_rank_one_update}
\end{aligned}
\end{equation}
which shows that a single pattern replacement changes only one column of $\mathbf H_{\rm eff}$.

\subsubsection{Cyclic Candidate-Testing}
Let $ \left(\overline{\mathbf R}, \mathbf Z^{[0]}, \mathbf F_{\rm RF}^{[0]}, \mathbf F_{\rm BB}^{[0]} \right)$ denote the feasible initial input configuration obtained from the preceding outer AO iteration. 
At iteration \(p\), construct the feasible testing set: $ \mathcal T^{[p]}
 \triangleq
 \Big\{
 \mathbf Z_{n_p\leftarrow s}^{[p-1]}:
 s\in\mathcal S,~T_{\rm d}
 \big(
 \mathbf Z_{n_p\leftarrow s}^{[p-1]}
 \big)
 \geq\epsilon_T
 \Big\}$. Since \(s=s_{n_p}^{[p-1]}\) reproduces the current pattern configuration, the current point satisfies $ \mathbf Z^{[p-1]}\in\mathcal T^{[p]}$. Thus, \(\mathcal T^{[p]}\neq \emptyset\). 

For every \(\widetilde{\mathbf Z}\in\mathcal T^{[p]}\), $\mathbf H_{\rm eff}$ 
is first generated using~\eqref{eq:candidate_channel_rank_one_update}. Thereafter, $(\mathbf F_{\rm BB}, \mathbf F_{\rm RF})$ are then reoptimized as $ \Big(
 \widehat{\mathbf F}_{\rm RF}^{[p]}
 (\widetilde{\mathbf Z}),
 \widehat{\mathbf F}_{\rm BB}^{[p]}
 (\widetilde{\mathbf Z})
 \Big)
 =
 \mathcal A_{\rm F}
 \Big(
 \overline{\mathbf R},
 \widetilde{\mathbf Z}
 \Big)$, where \(\mathcal A_{\rm F}\) denotes the framework in Section~\ref{subsec:dinkelbach_wmmse}. 
The corresponding locally optimized value is $ \widehat{\mathcal V}^{[p]}
 (\overline{\mathbf R}, \widetilde{\mathbf Z})
 \triangleq
 \eta_{\rm EE}
 \Big(
 \widehat{\mathbf F}_{\rm RF}^{[p]}
 (\widetilde{\mathbf Z}),
 \widehat{\mathbf F}_{\rm BB}^{[p]}
 (\widetilde{\mathbf Z}),
 \overline{\mathbf R}, 
 \widetilde{\mathbf Z}
 \Big)$, and the best configuration in $\mathcal T^{[p]}$ is selected as\footnote{Note that this compares at most \(S\) explicitly constructed candidates and is not exhaustive over the entire \(S^N\)-element patterns.}
\begin{equation}
 \widehat{\mathbf Z}^{[p]}(\overline{\mathbf R})
 \in
 \argmax_{
 \widetilde{\mathbf Z}\in\mathcal T^{[p]}
 }
 \widehat{\mathcal V}^{[p]}
 (\overline{\mathbf R}, \widetilde{\mathbf Z}).
 \label{eq:pattern_testing_set_selection}
\end{equation}
Let the EE at the beginning of iteration \(p\) be $ \eta_{\rm EE}^{[p-1]}
 \triangleq
 \eta_{\rm EE}
 \left(
 \mathbf F_{\rm RF}^{[p-1]},
 \mathbf F_{\rm BB}^{[p-1]},
 \overline{\mathbf R}, 
 \mathbf Z^{[p-1]}
 \right)$. 
From~\eqref{eq:dinkelbach_inner_ascent}-\eqref{eq:dinkelbach_parameter_monotonicity} and $\mathbf Z^{[p-1]}\in\mathcal T^{[p]}$: $\widehat{\mathcal V}^{[p]}
\left(
\overline{\mathbf R},
\mathbf Z^{[p-1]}
\right)
\geq
\eta_{\rm EE}^{[p-1]}$. Furthermore, by~\eqref{eq:pattern_testing_set_selection}:
\begin{equation}
\widehat{\mathcal V}^{[p]}
\left(
\overline{\mathbf R},
\widehat{\mathbf Z}^{[p]}
\right)
=
\max_{\widetilde{\mathbf Z}\in\mathcal T^{[p]}}
\widehat{\mathcal V}^{[p]}
\left(
\overline{\mathbf R},
\widetilde{\mathbf Z}
\right)
\geq
\widehat{\mathcal V}^{[p]}
\left(
\overline{\mathbf R},
\mathbf Z^{[p-1]}
\right)
\geq
\eta_{\rm EE}^{[p-1]}.
\label{eq:selected_pattern_candidate_monotonicity}
\end{equation}
Therefore, the selected $\widehat{\mathbf Z}^{[p]}$ together with its reoptimized $(\widehat{\mathbf F}_{\rm RF}^{[p]}
 (\widetilde{\mathbf Z}),
 \widehat{\mathbf F}_{\rm BB}^{[p]}
 (\widetilde{\mathbf Z}))$ cannot decrease the EE since by~\eqref{eq:pattern_testing_set_selection}: $ \eta_{\rm EE}^{[p]}\leftarrow
 \widehat{\mathcal V}^{[p]}
\left(
\overline{\mathbf R},
\widehat{\mathbf Z}^{[p]}
\right)$.
Hence for fixed $\overline{\mathbf R}$:
\begin{equation}
 \left(
 \mathbf Z^{[p]},
 \mathbf F_{\rm RF}^{[p]},
 \mathbf F_{\rm BB}^{[p]}
 \right)=
 \left(
 \widehat{\mathbf Z}^{[p]},
 \widehat{\mathbf F}_{\rm RF}^{[p]}
   (\widehat{\mathbf Z}^{[p]}),
 \widehat{\mathbf F}_{\rm BB}^{[p]}
   (\widehat{\mathbf Z}^{[p]})
 \right).
 \label{eq:joint_pattern_precoder_update}
\end{equation}

Define the pattern-change indicator $ b_{\rm pat}^{[p]} \triangleq
 \mathbf1_{\{
 \mathbf Z^{[p]}\neq\mathbf Z^{[p-1]}
 \}}$. 
 The procedure terminates when no pattern state has changed during the latest complete sweep, $ \sum_{i=p-N+1}^{p}
 b_{\rm pat}^{[i]}
 =0~ (p\geq N)$, and $ 
 |\eta_{\rm EE}^{[p]}-\eta_{\rm EE}^{[p-N]}|
 \leq\epsilon_{\rm COD}$, for given tolerance $\epsilon_{\rm COD}$.

\subsection{WMMSE-SCA-Based Optimization of $\mathbf R$}\label{optr}
\subsubsection{Quadratic Transform of $\eta_{\rm EE}$}
We then optimize $\mathbf R$ while holding others. Let
$(\overline{\mathbf F}_{\rm RF}, \overline{\mathbf F}_{\rm BB})$ and $\overline{\mathbf Z}\triangleq [\mathbf e_{\overline s_1}\cdots\mathbf e_{\overline s_N}]$ denote the fixed hybrid precoders and pattern, and define
\begin{equation}
\begin{aligned}
& \overline{\mathbf P}
 \triangleq
 \overline{\mathbf F}_{\rm RF}
 \overline{\mathbf F}_{\rm BB}
 =
 [\overline{\mathbf p}_1\cdots\overline{\mathbf p}_K],~\overline P_{\rm data}
 \triangleq
 \frac{\|\overline{\mathbf F}_{\rm BB}\|_{ F}^{2}}
 {\eta_{\rm PA}}+P_{\rm circ},\\
& \overline T_{\rm pat}
 \triangleq T_{\rm pat}(\overline{\mathbf Z}),~
 \overline E_{\rm pat}
 \triangleq E_{\rm pat}(\overline{\mathbf Z}).
\end{aligned}
\label{eq:fixed_variables_position_block}
\end{equation}
Let $\mathbf r\triangleq\operatorname{vec}(\mathbf R)$. Under Theorem~\ref{prop:optimal_speed}, define the $\mathbf r$-dependent:
\begin{equation}
\begin{aligned}
& \widehat T_{\rm mov}(\mathbf r)
 \triangleq
 T_{\rm mov}\left(
 \mathbf R,\widehat{\mathbf v}(\mathbf R)
 \right),
 E_{\rm mov}(\mathbf r)
 \triangleq
 E_{\rm mov}\left(
 \mathbf R,\widehat{\mathbf v}(\mathbf R)
 \right),\\
 &E_{\rm blk}(\mathbf r)
 \triangleq
 E_{\rm mov}(\mathbf r)
 +\overline E_{\rm pat}
 +T_{\rm d}(\mathbf r)\overline P_{\rm data},\\
 & T_{\rm d}(\mathbf r)
 \triangleq
 T_c
 -\widehat T_{\rm mov}(\mathbf r)
 -\overline T_{\rm pat},\eta_{\rm R}(\mathbf r)
 \triangleq
 \frac{
 WT_{\rm d}(\mathbf r)R_{\rm w}(\mathbf r)}
 {E_{\rm blk}(\mathbf r)}.
 \label{eq:reduced_position_ee}
\end{aligned}
\end{equation}
Here, $R_{\rm w}(\mathbf r)$ denotes $R_{\rm w}$ evaluated at $\mathbf R=\operatorname{unvec}(\mathbf r)$. 

To start, the quadratic transform gives~\cite{fracp}
\begin{equation}
\eta_{\rm EE}= \frac{A(\mathbf r)}{B(\mathbf r)}
 =\max_{\xi\geq0}
 \left[
 2\xi\sqrt{A(\mathbf r)}
 -\xi^2B(\mathbf r)
 \right],
 \label{eq:scalar_quadratic_transform}
\end{equation}
wherein $ A(\mathbf r)
 =
 WT_{\rm d}(\mathbf r)R_{\rm w}(\mathbf r)$ and $ B(\mathbf r)
 =
 E_{\rm blk}(\mathbf r)$, and the optimal $\xi^\star(\mathbf r)$ is $ \xi^\star(\mathbf r)
 =
 \frac{
 \sqrt{
 WT_{\rm d}(\mathbf r)
 R_{\rm w}(\mathbf r)
 }
 }{
 E_{\rm blk}(\mathbf r)
 }$. 
Thus, at iteration \(q\) with corresponding $\mathbf r^{(q)}$, 
we set $ \xi^{(q)}
 =
 \frac{
 \sqrt{
 WT_{\rm d}(\mathbf r^{(q)})
 R_{\rm w}(\mathbf r^{(q)})
 }
 }{
 E_{\rm blk}(\mathbf r^{(q)})
 }$.
 
\subsubsection{WMMSE Representation and $\mathbf r$-Dependent MSE}
For fixed $\overline s_n$, 
the coefficient of stream \(j\) observed by user \(k\) is $ a_{k,j}(\mathbf r)
 \triangleq
 \mathbf h_k^*(\mathbf R,\overline{\mathbf Z})
 \overline{\mathbf p}_j
 =
 \sum_{n=1}^{N}
 \overline p_{n,j}
 \chi_{k,n}(\mathbf r_n)$ with $\overline p_{n,j}\triangleq[\overline{\mathbf p}_j]_n$ and $\chi_{k,n}\triangleq \chi_{k,n, \overline s_n}$. Same with Section~\ref{subsec:dinkelbach_wmmse}, for equalizer \(\tilde{u}_k\in\mathbb C\), the MSE of user \(k\) is $ e_k(\mathbf r,\tilde{u}_k)
 =
 |\tilde{u}_k|^2
 \left(
 \sum_{j=1}^{K}
 |a_{k,j}(\mathbf r)|^2
 +\sigma_k^2
 \right)
 -
 2\Re\left\{
 \tilde{u}_k^*a_{k,k}(\mathbf r)
 \right\}
 +1$, and the sum-SE satisfies~\cite{Zheng2025TriHybrid}: 
\begin{equation}
\begin{aligned}
 R_{\rm w}(\mathbf r)
 =
 \frac{1}{\ln2}
 \max_{
 \{\tilde{u}_k,\tilde{\varpi}_k>0\}
 }
 \sum_{k=1}^{K}
 \left(
 \ln\tilde{\varpi}_k
 -\tilde{\varpi}_ke_k(\mathbf r,\tilde{u}_k)
 +1
 \right),
 \label{eq:position_wmmse_identity}
\end{aligned}
\end{equation}
where \(\tilde{\varpi}_k>0\) is the auxiliary MSE weight, and at \(\mathbf r^{(q)}\), the optimal $\tilde{u}_k$ and $\tilde{\varpi}_k$ are respectively $ \tilde{u}_k^{(q)}
 =
 \frac{
 a_{k,k}(\mathbf r^{(q)})
 }{
 \sum_{j=1}^{K}
 |a_{k,j}(\mathbf r^{(q)})|^2
 +\sigma_k^2
 }$ and  $ \tilde{\varpi}_k^{(q)}
 =
 \frac{1}{
 e_k(\mathbf r^{(q)},\tilde{u}_k^{(q)})
 }$. 
 However, since \(e_k(\mathbf r,\tilde{u}_k^{(q)})\) remains nonconvex in \(\mathbf r\), we construct its convex upper-approximation.

\subsubsection{Convex MSE Majorization}
Using  $ \mathbf g_{k,n}(\mathbf r_n)
 \triangleq
 \nabla_{\mathbf r_n}
 \chi_{k,n}(\mathbf r_n)
 =
 - j
 \sqrt{\frac{\beta_k}{L_k}}
 \sum_{\ell=1}^{L_k}
 \alpha_{k,\ell}
 \overline\psi_{\overline s_n}
 (\boldsymbol\Omega_{k,\ell})
 \boldsymbol\kappa_{k,\ell}
 e^{-j
 \boldsymbol\kappa_{k,\ell}^{\mathrm T}\mathbf r_n}$:
\begin{equation}
\begin{aligned}
 \nabla_{\mathbf r_n}
 e_k(\mathbf r,\tilde{u}_k)
 =
 &2|\tilde{u}_k|^2
 \sum_{j=1}^{K}
 \Re\left\{
 a_{k,j}^*(\mathbf r)
 \overline p_{n,j}
 \mathbf g_{k,n}(\mathbf r_n)
 \right\}
 \\
 &-
 2\Re\left\{
 \tilde{u}_k^*
 \overline p_{n,k}
 \mathbf g_{k,n}(\mathbf r_n)
 \right\},
 \label{eq:position_mse_gradient}
\end{aligned}
\end{equation}
and stacking~\eqref{eq:position_mse_gradient} over \(n\) gives \(\nabla_{\mathbf r}e_k\). Since $\{\mathcal C_n\}$ are compact and the field-response functions are twice continuously differentiable, the gradient of $e_k$ is Lipschitz continuous. Hence, for a sufficiently large curvature \(L_{e,k}^{(q)}>0\)~\cite{Scutari2017SCA}:\footnote{In general, the curvature is initialized as \(L_{e,k}^{(q,0)}=\|\nabla_{\mathbf r}^{2} e_k(\mathbf r^{(q)},\tilde{u}_k^{(q)})\|_2\). It is then multiplied by \(c_L>1\) until~\eqref{eq:mse_quadratic_majorizer} holds at the candidate point~\cite{Scutari2017SCA}.}
\begin{equation}
\begin{aligned}
 e_k(\mathbf r,\tilde{u}_k^{(q)})
 \leq&
 e_k(\mathbf r^{(q)},\tilde{u}_k^{(q)})
 +
 \nabla_{\mathbf r}
 e_k(\mathbf r^{(q)},\tilde{u}_k^{(q)})^{\mathrm T}
 (\mathbf r-\mathbf r^{(q)})
 \\
 &+
 \frac{L_{e,k}^{(q)}}{2}
 \|\mathbf r-\mathbf r^{(q)}\|_2^2
 \triangleq
 \overline e_k
 (\mathbf r;\mathbf r^{(q)}).
 \label{eq:mse_quadratic_majorizer}
\end{aligned}
\end{equation}
Herein, 
\(\overline e_k(\mathbf r;\mathbf r^{(q)})\) is convex in \(\mathbf r\), upper-bounds $e_k$, and satisfies $ \overline e_k
 (\mathbf r^{(q)};\mathbf r^{(q)})
 =
 e_k(\mathbf r^{(q)},\tilde{u}_k^{(q)})$ and $\nabla_{\mathbf r}
 \overline e_k
 (\mathbf r^{(q)};\mathbf r^{(q)})
 =
 \nabla_{\mathbf r}
 e_k(\mathbf r^{(q)},\tilde{u}_k^{(q)})$. Substituting~\eqref{eq:mse_quadratic_majorizer} into~\eqref{eq:position_wmmse_identity} yields
\begin{equation}
\begin{aligned}
 \underline R_{\rm w}
 (\mathbf r;\mathbf r^{(q)})
 \triangleq
 \frac{1}{\ln2}
 \sum_{k=1}^{K}
 \Big[
 &\ln\tilde{\varpi}_k^{(q)}
 -
 \tilde{\varpi}_k^{(q)}
 \overline e_k
 (\mathbf r;\mathbf r^{(q)})
 +1
 \Big].
 \label{eq:concave_rate_minorizer}
\end{aligned}
\end{equation}
Since
\(-\overline e_k\) is concave,
\(\underline R_{\rm w}\) is concave in \(\mathbf r\). Moreover, $ \underline R_{\rm w}
 (\mathbf r;\mathbf r^{(q)})
 \leq
 R_{\rm w}(\mathbf r)~(\forall \mathbf r)$, 
while equality and first-order consistency hold at \(\mathbf r=\mathbf r^{(q)}\).

\subsubsection{Convex Majorization of $ E_{\rm blk}$}
Although \(T_{\rm d}(\mathbf r)\) is concave, the denominator term
\(T_{\rm d}(\mathbf r)\overline P_{\rm data}\) must be upper-bounded to construct a valid minorizer of the transformed EE objective. Define $ \mathcal I^{(q)}
 \triangleq
 \argmax_{\substack{n\in\mathcal N\\a\in\mathcal A}}
 \frac{
 |r_{n,a}^{(q)}-r_{n,a}^0|
 }{
 v_{n,a}^{\max}
 }$ the active set of $(n, a)$ that determine the overall mechanical reconfiguration time at \(\mathbf r^{(q)}\). A subgradient of
\(\widehat T_{\rm mov}\) at \(\mathbf r^{(q)}\)~\cite{boyd}: 
\begin{equation}
\label{subg}
 \mathbf g_{\rm mov}^{(q)}
 =
 \sum_{(n,a)\in\mathcal I^{(q)}}
 \vartheta_{n,a}^{(q)}
 \frac{
 \operatorname{sgn}
 (r_{n,a}^{(q)}-r_{n,a}^0)
 }{
 v_{n,a}^{\max}
 }
 \mathbf e_{n,a},
 \end{equation}
  where $ \vartheta_{n,a}^{(q)}\geq0,~
 \sum_{(n,a)\in\mathcal I^{(q)}}
 \vartheta_{n,a}^{(q)}=1$, and \(\mathbf e_{n,a}\in\mathbb R^{2N}\) is the canonical basis associated with \((n,a)\). Accordingly, $ \mathbf g_{\rm d}^{(q)} \triangleq -\mathbf g_{\rm mov}^{(q)}$ is a supergradient of the concave function \(T_{\rm d}\), and its first-order affine upper-bound is therefore
\begin{equation}
 T_{\rm d}(\mathbf r)
 \leq
 T_{\rm d}(\mathbf r^{(q)})
 +
 (\mathbf g_{\rm d}^{(q)})^{\mathrm T}
 (\mathbf r-\mathbf r^{(q)})
 \triangleq
 \overline T_{\rm d}
 (\mathbf r;\mathbf r^{(q)}).
 \label{eq:data_time_affine_upper_bound}
\end{equation}
The inequality direction follows from the concavity of \(T_{\rm d}\).

 Within every differentiable operating region, the gradient of \(E_{\rm mov}\) is Lipschitz continuous. Thus, with $ \mathbf g_E^{(q)}
 \triangleq
 \nabla_{\mathbf r}
 E_{\rm mov}(\mathbf r^{(q)})$ and for a sufficiently large
\(L_E^{(q)}>0\),
\begin{equation}
\begin{aligned}
 E_{\rm mov}(\mathbf r)
 \leq&
\underbrace{ E_{\rm mov}(\mathbf r^{(q)})
 +
 (\mathbf g_E^{(q)})^{\mathrm T}
 (\mathbf r-\mathbf r^{(q)})
 +
 \frac{L_E^{(q)}}{2}
 \|\mathbf r-\mathbf r^{(q)}\|_2^2}
 _{\triangleq \overline E_{\rm mov}
 (\mathbf r;\mathbf r^{(q)})}.
 \label{eq:movement_energy_quadratic_majorizer}
\end{aligned}
\end{equation}
Combining~\eqref{eq:data_time_affine_upper_bound} and~\eqref{eq:movement_energy_quadratic_majorizer} gives the upper-bound of $E_{\rm blk}$:
\begin{equation}
 \overline E_{\rm blk}
 (\mathbf r;\mathbf r^{(q)})
 \triangleq
 \overline E_{\rm mov}
 (\mathbf r;\mathbf r^{(q)})
 +\overline E_{\rm pat}
 +
 \overline P_{\rm data}
 \overline T_{\rm d}
 (\mathbf r;\mathbf r^{(q)}),
 \label{eq:convex_block_energy_majorizer}
\end{equation}
which is convex, satisfies $ \overline E_{\rm blk}
 (\mathbf r;\mathbf r^{(q)})
 \geq
 E_{\rm blk}(\mathbf r)$ and is tight and first-order consistent at
\(\mathbf r^{(q)}\).

\subsubsection{Successive Inner Approximation of $\|\mathbf r_n-\mathbf r_m\|_2^2$}
For each \(n<m\), define $ \boldsymbol q_{n,m}^{(q)}
 \triangleq
 \mathbf r_n^{(q)}-\mathbf r_m^{(q)}$. Since \(\|\mathbf r_n-\mathbf r_m\|_2^2\) is convex, its first-order lower-bound at current position is $ \|\mathbf r_n-\mathbf r_m\|_2^2
 \geq
 \|\boldsymbol q_{n,m}^{(q)}\|_2^2
+
 2(\boldsymbol q_{n,m}^{(q)})^{\mathrm T}
 \left[
 (\mathbf r_n-\mathbf r_m)
 -\boldsymbol q_{n,m}^{(q)}
 \right]$. Consequently, the affine constraint
\begin{equation}
\begin{aligned}
 \|\boldsymbol q_{n,m}^{(q)}\|_2^2
 +
 2(\boldsymbol q_{n,m}^{(q)})^{\mathrm T}
 \left[
 (\mathbf r_n-\mathbf r_m)
 -\boldsymbol q_{n,m}^{(q)}
 \right]
 \geq d_{\min}^2
 \label{eq:joint_spacing_inner_constraint}
\end{aligned}
\end{equation}
is sufficient for satisfying the original spacing constraint, which is also tight at
\((\mathbf r_n^{(q)},\mathbf r_m^{(q)})\).

Define the convex inner position set $ \mathcal D_{\rm R}^{(q)}
 \triangleq
 \Big\{
 \mathbf r:
 \mathbf r_n\in\mathcal C_n~(\forall n),~\eqref{eq:joint_spacing_inner_constraint}~(\forall n<m),~T_{\rm d}(\mathbf r)\geq\epsilon_T, \|\mathbf r-\mathbf r^{(q)}\|_2
 \leq\Delta^{(q)}
 \Big\}$, where \(\Delta^{(q)}>0\) is an optional trust-region radius used to ensure the accuracy of the local quadratic majorizers, and $T_{\rm d}(\mathbf r)\geq\epsilon_T$, a superlevel set of the concave \(T_{\rm d}\), is convex.

\subsubsection{SCA Transformation}
For fixed \(\xi^{(q)}\), define the transformed surrogate: $ \underline{\mathcal Q}_{\rm R}
 (\mathbf r;\mathbf r^{(q)})
 \triangleq
 2\xi^{(q)}
 \sqrt{
 WT_{\rm d}(\mathbf r)
 \underline R_{\rm w}
 (\mathbf r;\mathbf r^{(q)})
 }
-
 (\xi^{(q)})^2
 \overline E_{\rm blk}
 (\mathbf r;\mathbf r^{(q)})
 -
 \frac{\rho_{\rm R}}{2}
 \|\mathbf r-\mathbf r^{(q)}\|_2^2$, where \(\rho_{\rm R}>0\) is a proximal regularization coefficient. Herein, since \(T_{\rm d}\) and
\(\underline R_{\rm w}\) are nonnegative concave functions, their geometric mean $ \sqrt{
 T_{\rm d}(\mathbf r)
 \underline R_{\rm w}
 (\mathbf r;\mathbf r^{(q)})
 }$ is concave. Moreover,
\(\overline E_{\rm blk}\) is convex. Hence,
\(\underline{\mathcal Q}_{\rm R}\) is concave in
\(\mathbf r\).

To obtain a solver-ready formulation, introduce auxiliary variables \(x\geq0\), \(y\geq0\), and \(\zeta\geq0\), representing a lower-bound on data time and rate, and geometric-mean, respectively. The \(q\)th SCA subproblem is
\begin{equation}
\begin{aligned}
& \max_{
 \mathbf r,x,y,\zeta\ge0
 }
 ~
 2\xi^{(q)}\zeta
 -
 (\xi^{(q)})^2
 \overline E_{\rm blk}
 (\mathbf r;\mathbf r^{(q)})
 -
 \frac{\rho_{\rm R}}{2}
 \|\mathbf r-\mathbf r^{(q)}\|_2^2
\\
 &\mathrm{s.t.}~
 \mathbf r\in\mathcal D_{\rm R}^{(q)},~\epsilon_T\leq x
 \leq T_{\rm d}(\mathbf r),~\epsilon_R\leq y
 \leq
 \underline R_{\rm w}
 (\mathbf r;\mathbf r^{(q)}),
\\
 &~~~~\left\|
 \begin{bmatrix}
 2\zeta/\sqrt W\\
 x-y
 \end{bmatrix}
 \right\|_2
 \leq x+y.
 \label{prob:position_rotated_soc}
\end{aligned}
\end{equation}
Here, \(\epsilon_R>0\) ensures that $\underline R_{\rm w}$ 
remains positive. The last constraint of~\eqref{prob:position_rotated_soc} is equivalent to $ \zeta^2\leq Wxy$, and is a rotated second-order-cone constraint. Therefore,~\eqref{prob:position_rotated_soc} is a convex optimization problem that can be solved by CVX~\cite{boyd}.

\subsubsection{Minorization Property and Monotonicity}
From 
$ \underline R_{\rm w}
 (\mathbf r;\mathbf r^{(q)})
 \leq R_{\rm w}(\mathbf r),
 ~
 \overline E_{\rm blk}
 (\mathbf r;\mathbf r^{(q)})
 \geq E_{\rm blk}(\mathbf r)$:
\begin{equation}
 \underline{\mathcal Q}_{\rm R}
 (\mathbf r;\mathbf r^{(q)})
 \leq
 2\xi^{(q)}
 \sqrt{
 WT_{\rm d}(\mathbf r)
 R_{\rm w}(\mathbf r)
 }
 -
 (\xi^{(q)})^2
 E_{\rm blk}(\mathbf r)
 \leq
 \eta_{\rm R}(\mathbf r),
 \label{eq:position_global_minorizer}
\end{equation}
where the second inequality follows from the optimality condition of~\eqref{eq:scalar_quadratic_transform}. Furthermore, $ \underline{\mathcal Q}_{\rm R} (\mathbf r^{(q)};\mathbf r^{(q)}) = \eta_{\rm R}(\mathbf r^{(q)})$. Let \({\mathbf r}^{(q+1)}\) be the optimizer of~\eqref{prob:position_rotated_soc}. Since \(\mathbf r^{(q)}\in\mathcal D_{\rm R}^{(q)}\):
\begin{equation}
 \underline{\mathcal Q}_{\rm R}
 ({\mathbf r}^{(q+1)};
  \mathbf r^{(q)})
 \geq
 \underline{\mathcal Q}_{\rm R}
 (\mathbf r^{(q)};
  \mathbf r^{(q)}).
 \label{eq:surrogate_improvement}
\end{equation}
Combining~\eqref{eq:position_global_minorizer}-\eqref{eq:surrogate_improvement} yields
\begin{equation}
 \eta_{\rm R}
 ({\mathbf r}^{(q+1)})
 \geq
 \eta_{\rm R}
 (\mathbf r^{(q)}).
 \label{eq:position_monotonicity}
\end{equation}
Thus, every valid SCA update is guaranteed to be nondecreasing in the original EE.
The SCA procedure terminates when $ 
 |\eta_{\rm R}(\mathbf r^{(q+1)})
 -\eta_{\rm R}(\mathbf r^{(q)})|
 \leq\epsilon_{\rm SCA}$, for given tolerance $\epsilon_{\rm SCA}$.

\begin{remark}
\label{atd}
At an iterate corresponding to a zero displacement or an axis tie, \(E_{\rm mov}(\mathbf r)\) may be nondifferentiable. In this case, let \(E_{{\rm mov},\tau}(\mathbf r)\) denote a continuously differentiable approximation of \(E_{\rm mov}(\mathbf r)\), obtained by replacing its absolute-value and maximum operators with smooth approximations $\phi_{\tau}(x)$ and $m_{\tau}(\mathbf z)$ parameterized by \(\tau>0\), respectively~\cite{boyd}: $\phi_{\tau}(x)=\sqrt{x^2+\tau^2}$ and $m_{\tau}(\mathbf z)
=\tau\ln\sum_i e^{z_i/\tau}$ so that $\lim_{\tau\leftarrow0^+}E_{{\rm mov},\tau}(\mathbf r)
=
E_{\rm mov}(\mathbf r)$. For a fixed \(\tau_q>0\), a candidate \({\mathbf r}^{(q+1)}\) is generated using the quadratic model
\begin{equation}
\begin{aligned}
\overline E_{{\rm mov},\tau_q}
(\mathbf r;\mathbf r^{(q)})
\triangleq {}&
E_{{\rm mov},\tau_q}(\mathbf r^{(q)})
+
\nabla E_{{\rm mov},\tau_q}(\mathbf r^{(q)})^{\mathrm T}
(\mathbf r-\mathbf r^{(q)})
\\
&+
\frac{L_{E,\tau_q}^{(q)}}{2}
\|\mathbf r-\mathbf r^{(q)}\|_2^2.
\label{tsm}
\end{aligned}
\end{equation}
Since~\eqref{tsm} does not necessarily majorize the original nonsmooth \(E_{\rm mov}\), the resulting ${\mathbf r}^{(q+1)}$ at the end of the optimization is accepted only after the evaluation:
\begin{equation}
\label{oee3s}
\mathbf r^{(q+1)}
\leftarrow
\begin{cases}
{\mathbf r}^{(q+1)}
&
(\substack{{\mathbf r}^{(q+1)}\text{ is feasible and}\\
\eta_{\rm R}({\mathbf r}^{(q+1)})\ge\eta_{\rm R}(\mathbf r^{(q)})}),
\\
\mathbf r^{(q)}
&(\text{otherwise}).
\end{cases}
\end{equation}
Upon rejection, \(\tau_q\) 
may be reduced by fixed portion and the surrogate problem may be re-solved. Consequently, $\eta_{\rm R}(\mathbf r^{(q+1)}) \ge \eta_{\rm R}(\mathbf r^{(q)})$ holds. 
The reliability of the $\mathbf R$-convergence will be numerically verified in Section~\ref{subsec:reliability_optimality}.
\end{remark}

\subsection{Overall AO Framework}
\label{subsec:overall_ao}

\begin{algorithm}[t]
\caption{Proposed AO Framework for EE Maximization}
\label{alg:overall_ao}
\begin{algorithmic}[1]
\Require Feasible
\(\mathcal X^{(0)}
=
\{\mathbf F_{\rm BB}^{(0)},\mathbf F_{\rm RF}^{(0)},
\mathbf R^{(0)},\mathbf Z^{(0)}\}\)
and tolerances \(\{\epsilon_{(\cdot)}\}\)

\State \(\mathbf F^{(0)}
\triangleq
(\mathbf F_{\rm BB}^{(0)},\mathbf F_{\rm RF}^{(0)})\), \(t\gets0\), and evaluate \(\eta_{\rm EE}^{(0)}\).

\Repeat
    \State Obtain $    \overline{\mathbf F}^{(t)}
    \gets
    \mathcal A_{\rm F}
    \big(
    \mathbf R^{(t)},\mathbf Z^{(t)};
    \mathbf F^{(t)}
    \big)$.  
    \State Obtain
    \((\mathbf Z^{(t+1)},\mathbf F^{(t+1)})\) 
     and \(\mathbf R^{(t+1)}\)


    \State     \(\mathbf v^{(t+1)}
    \gets\widehat{\mathbf v}(\mathbf R^{(t+1)})\),
    evaluate \(\eta_{\rm EE}^{(t+1)}\), and \(t\gets t+1\).

\Until{
\(
|\eta_{\rm EE}^{(t)}-\eta_{\rm EE}^{(t-1)}|
\leq\epsilon_{\rm AO}
\)}

\State \textbf{Return}
\(\mathcal X^\star\gets\mathcal X^{(t)}\) and
\(\mathbf v^\star
\gets\widehat{\mathbf v}(\mathbf R^\star)\).
\end{algorithmic}
\end{algorithm}
By combining Section~\ref{subsec:optimal_speed}-\ref{optr}, we propose the AO framework for EE maximization with details in Algorithm~\ref{alg:overall_ao}, where the single-$\mathbf F$ notation corresponds to $(\mathbf F_{\rm BB}, \mathbf F_{\rm RF})$.\footnote{Note that \((\mathbf R^0,\mathbf Z^0)\), from which the reconfiguration overhead is measured, is distinguished from the iterates \((\mathbf R^{(t)},\mathbf Z^{(t)})\), especially for $t=0$.} 
For deterministic initialization \(\mathcal X^{(0)}\), 
we assume that $S\ge2$ and $ T_c-\epsilon_T>\tau_{\rm sw}$ with sufficiently small $\epsilon_T$.

We first select a prescribed anchor \(\hat{\mathbf r}_n(\neq\mathbf r_n^0)\in\mathcal C_n\) and define $\Delta\mathbf r_n
\triangleq
\hat{\mathbf r}_n-\mathbf r_n^0$ and $\Delta\mathbf r_{n,m}
\triangleq
\Delta\mathbf r_n-\Delta\mathbf r_m$. 
To ensure that $\mathbf Z^{(0)}\neq\mathbf Z^0$, select a prescribed MPRA \(\bar n\) and cyclically change its state: $s_{\bar n}^{(0)}
=
1+\operatorname{mod}(s_{\bar n}^{0},S),
~
s_n^{(0)}=s_n^0
~(n\neq\bar n)$, where \(\mathbf z_n^0=\mathbf e_{s_n^0}\), and set $\mathbf Z^{(0)}
=
[\mathbf e_{s_1^{(0)}}\cdots
 \mathbf e_{s_N^{(0)}}]$. Thus, provided that \(S\geq2\), $\mathbf Z^{(0)}\neq\mathbf Z^0,
~
T_{\rm pat}(\mathbf Z^{(0)})=\tau_{\rm sw}$.

Now define $\overline T_{\rm mov}
\triangleq
\max_{\substack{n\in\mathcal N\\a\in\mathcal A}}
\frac{
|[\Delta\mathbf r_n]_a|
}{
v_{n,a}^{\max}
}$ and the time-feasible scaling bound: $\alpha_T
\triangleq
\begin{cases}
\frac{T_c-\epsilon_T-\tau_{\rm sw}}
{\overline T_{\rm mov}}
&(\overline T_{\rm mov}>0)\\
+\infty
&(\overline T_{\rm mov}=0)
\end{cases}(>0)$. A conservative spacing-feasible scaling bound is also $\alpha_{\rm sp}
\triangleq
\min_{\substack{n<m:\\
\|\Delta\mathbf r_{n,m}\|_2>0}}
\frac{
\|\mathbf r_n^0-\mathbf r_m^0\|_2-d_{\min}
}{
\|\Delta\mathbf r_{n,m}\|_2
}(>0)$, where the minimum over an empty set is interpreted as \(+\infty\).
For a prescribed \(\chi\in(0,1)\), set $\alpha_0
=
\chi
\min\{1,\alpha_T,\alpha_{\rm sp}\}(>0)$ and $\mathbf r_n^{(0)}
=
\mathbf r_n^0+\alpha_0\Delta\mathbf r_n$. Since every \(\mathcal C_n\) is convex,
\(\mathbf r_n^{(0)}\in\mathcal C_n\). Moreover, 
\begin{equation}
\begin{aligned}
\|\mathbf r_n^{(0)}-\mathbf r_m^{(0)}\|_2
&=
\|
\mathbf r_n^0-\mathbf r_m^0
+\alpha_0\Delta\mathbf r_{n,m}
\|_2
\\
&\geq
\|\mathbf r_n^0-\mathbf r_m^0\|_2
-\alpha_0\|\Delta\mathbf r_{n,m}\|_2
\geq d_{\min},
\label{eq:initial_spacing_guarantee}
\end{aligned}
\end{equation}
obtained by the reverse triangle inequality. Furthermore,
\begin{align}
\widehat T_{\rm mov}(\mathbf R^{(0)})
+
T_{\rm pat}(\mathbf Z^{(0)})
&=
\alpha_0\overline T_{\rm mov}
+\tau_{\rm sw}
\leq
T_c-\epsilon_T.
\label{eq:initial_time_guarantee}
\end{align}
Hence $\mathbf R^{(0)}\neq\mathbf R^0$ and $\mathbf Z^{(0)}\neq\mathbf Z^0$, and the initialized position and EM-domain variables are feasible.

Furthermore, $\mathbf F_{\rm RF}^{(0)}$ is deterministically initialized as $\mathbf F_{\rm RF}^{(0)}
=
\operatorname{blkdiag}
\left(
\boldsymbol\phi_1^{(0)},\cdots,
\boldsymbol\phi_{N_{\rm RF}}^{(0)}
\right)$ with $[\boldsymbol\phi_r^{(0)}]_m
=
\frac{e^{j\theta_{r,m}^{(0)}}}{\sqrt M}$, where we set $\theta_{r,m}^{(0)}
=
\frac{2\pi(m-1)}{M}$. Finally, since \(N_{\rm RF}\geq K\): $\mathbf F_{\rm BB}^{(0)}
=
\sqrt{\frac{P_{\rm max}}{K}}
\begin{bmatrix}
\mathbf I_K\\
\mathbf0_{(N_{\rm RF}-K)\times K}
\end{bmatrix}$, which satisfies $\|\mathbf F_{\rm BB}^{(0)}\|_{F}^2\leq P_{\max}$. Therefore, $\left\{
\mathbf F_{\rm BB}^{(0)},
\mathbf F_{\rm RF}^{(0)},
\mathbf R^{(0)},
\mathbf Z^{(0)}
\right\}$ constitutes a deterministic feasible initialization.

The monotonicity and convergence of EE is given by following theorem:
\begin{theorem}
\label{prop:overall_ao_convergence}
\(\{\eta_{\rm EE}^{(t)}\}\) is nondecreasing and converges.
\end{theorem}

\begin{proof}
Let \(\eta_{\rm F}^{(t)}\) and \(\eta_{\rm Z}^{(t)}\) denote the original EEs after the update of $(\mathbf F_{\rm BB}, \mathbf F_{\rm RF})$ and $\mathbf Z$, respectively, during outer iteration \(t\). Since~\eqref{eq:dinkelbach_parameter_monotonicity} implies that the update of $(\mathbf F_{\rm BB}, \mathbf F_{\rm RF})$ does not decrease the original EE: $\eta_{\rm F}^{(t)}
\geq
\eta_{\rm EE}^{(t)}$. In $\mathbf Z$-block, by~\eqref{eq:selected_pattern_candidate_monotonicity}-\eqref{eq:joint_pattern_precoder_update}: $\eta_{\rm Z}^{(t)}
\geq
\eta_{\rm F}^{(t)}$. For $\mathbf R$-block,~\eqref{eq:position_monotonicity} ensure that the update is nondecreasing in the original EE. Even at a nondifferentiable iterate,~\eqref{oee3s} accepts the candidate only when it is feasible and does not decrease the original nonsmooth EE. Consequently, $\eta_{\rm EE}^{(t+1)}
\geq
\eta_{\rm Z}^{(t)}$. Combining the results yields $\eta_{\rm EE}^{(t)}
\leq
\eta_{\rm F}^{(t)}
\leq
\eta_{\rm Z}^{(t)}
\leq
\eta_{\rm EE}^{(t+1)}$. Thus, \(\{\eta_{\rm EE}^{(t)}\}\) is nondecreasing.

It remains to show that \(\{\eta_{\rm EE}^{(t)}\}\) is bounded above. From the data-time constraint and \(P_{\rm circ}>0\), $E_{\rm blk}
=
\widehat E_{\rm rec}
+
\widehat T_{\rm d}P_{\rm data}\geq
\widehat T_{\rm d}P_{\rm circ}
\geq
\epsilon_TP_{\rm circ}(>0)$. Moreover, \(\|\mathbf F_{\rm BB}\|_{ F}^2\leq P_{\max}\), $\mathcal F_{\rm RF}$ and $\{\mathcal C_n\}$ are compact, and the feasible set of $\mathbf Z$ is finite. 
Hence, there exists a finite constant \(\overline R_{\rm w}\) such that $R_{\rm w}\leq\overline R_{\rm w}$ over the feasible set. Furthermore, since \(0<\widehat T_{\rm d}\leq T_c\), it follows that $\eta_{\rm EE}^{(t)}
\leq
\frac{
WT_c\overline R_{\rm w}
}{
\epsilon_TP_{\rm circ}
}<\infty$. Therefore, \(\{\eta_{\rm EE}^{(t)}\}\) is nondecreasing and bounded above, and the theorem follows.
\end{proof}

\subsection{Computational Complexity}
\label{subsec:computational_complexity}
Let \(L_{\Sigma}\triangleq\sum_{k=1}^{K}L_k\) and \(B_{\rm R}\) be the maximum Armijo trials per RCG iteration. Let \(I_{\rm AO}\), \(I_{\rm D}\), \(I_{\mu}\), \(I_{\rm R}\), \(I_{\rm P}\), and \(I_{\rm SCA}\) denote, respectively, the numbers of outer AO iterations, Dinkelbach-WMMSE iterations for \(\mathcal A_{\rm F}\), bisection iterations per \(\mathbf F_{\rm BB}\)-update, RCG iterations per \(\mathbf F_{\rm RF}\)-update, individual pattern-coordinate updates per \(\mathbf Z\)-update, and SCA iterations per \(\mathbf R\)-update. 

\textbf{$(\mathbf F_{\rm BB}, \mathbf F_{\rm RF})$-Update}: For each $i$, computing the effective channels, equalizers, and MSE weights requires $\mathcal O \left(KN+K^{2}N_{\rm RF}\right)$ operations. Constructing \(\mathbf A\) and \(\mathbf B\) and evaluating~\eqref{eq:bb_closed_form} require $ \mathcal O \left(
N_{\rm RF}^{3}
+KN_{\rm RF}^{2}
+I_{\mu}KN_{\rm RF}
\right)$ operations, where the eigendecomposition of \(\mathbf A\) is reused
during the scalar bisection. For $\mathbf F_{\rm RF}$-update, explicitly constructing $ 
\mathbf C$ requires \(\mathcal O(K^{2}N^{2})\) operations. Given \(\mathbf C\), each RCG gradient evaluation requires \(\mathcal O(N^{2})\) operations~\cite{hbyx}, and the corresponding Armijo search requires at most \(\mathcal O(B_{\rm R}N^{2})\) operations~\cite{arpo}. The tangent-space projection, vector transport, and retraction require only \(\mathcal O(N)\) operations. Consequently, 
\begin{equation}
\begin{aligned}
\mathcal C_{\rm F}
=
\mathcal O\Big(&
I_{\rm D}\big[K^{2}N_{\rm RF}
+KN_{\rm RF}^{2}
+N_{\rm RF}^{3}
+I_{\mu}KN_{\rm RF}
\\
&+K^{2}N^{2}
+I_{\rm R}(1+B_{\rm R})N^{2}
\big]
\Big).
\label{eq:precoder_complexity}
\end{aligned}
\end{equation}

\textbf{$\mathbf Z$-Update}: For fixed \(\mathbf R\), constructing all \(\{\mathbf h_{n,s}\}_{n,s}\) requires \(\mathcal O(NSL_{\Sigma})\) operations. At each $p$, at most \(S\) candidates are tested with invoking \(\mathcal A_{\rm F}\). Therefore, 
\begin{equation}
\label{upz}
\mathcal C_{\rm Z}
=
\mathcal O \left(
NSL_{\Sigma}
+
I_{\rm P}S
\big[
\mathcal C_{\rm F}+K+N
\big]
\right),
\end{equation}
where the \(\mathcal O(K+N)\) term accounts for the rank-one channel update, feasibility test, and switching-overhead evaluation. 

\textbf{$\mathbf R$-Update}: For each $q$, evaluating the $\mathbf r$-dependent channels and constructing the WMMSE surrogates require $\mathcal O \left(NL_{\Sigma}+K^{2}N\right)$ operations. Problem~\eqref{prob:position_rotated_soc} has \(2N+3\) primary variables, and after standard epigraph and second-order cone reformulations, the number remains \(\widetilde n_{\rm R}=\mathcal O(N)\), whereas the number of scalar and conic constraints is $m_{\rm R}=\frac{N(N-1)}{2}+\mathcal O(N)=\mathcal O(N^{2})$. 
A generic primal-dual interior-point method has the conservative complexity bound~\cite[Chapter~11]{boyd}: 
$\mathcal C_{\rm IP}
=
\mathcal O \left(
N^{5}\log\frac{1}{\epsilon_{\rm IP}}
\right)$ with accuracy $\epsilon_{\rm IP}>0$. If at most \(B_{\rm S}\) convex subproblems are solved at each SCA iteration due to Remark~\ref{atd},\footnote{Hence \(B_{\rm S}=1\) when no re-solving is performed.} the $\mathbf R$-update requires
\begin{equation}
\mathcal C_{\rm R}
=
\mathcal O \left(
I_{\rm SCA}
\left[
NL_{\Sigma}
+K^{2}N
+B_{\rm S}\mathcal C_{\rm IP}
\right]
\right).
\label{eq:position_complexity}
\end{equation}

Hence, the overall computational complexity is
\begin{equation}
\begin{aligned}
\mathcal C_{\rm AO}
=
\mathcal O\Big(&
I_{\rm AO}\Big[
(1+I_{\rm P}S)\mathcal C_{\rm F}
+NSL_{\Sigma}
+I_{\rm P}S(K+N)
\\
&+
I_{\rm SCA}
\big(
NL_{\Sigma}
+K^{2}N
+B_{\rm S}\mathcal C_{\rm IP}
\big)
\Big]
\Big),
\label{eq:overall_ao_complexity}
\end{aligned}
\end{equation}
where the deterministic initialization in Section~\ref{subsec:overall_ao} requires \(\mathcal O(N^{2}+N_{\rm RF}K)\) operations and is thus negligible.

\section{Simulation Results}
\label{sec:simulation_results}
\subsection{Simulation Setup}
\label{subsec:simulation_setup}
We consider a multi-user downlink system with $f_c=3.55$~GHz and $\forall \sigma_k^2=-80$~dBm, in which a BS equipped with \(N=16\) MPRAs with $\{\mathcal C_n\}$ arranged as a \(4\times4\) uniform planar array (UPA) form, serves \(K=4\) single-antenna users using \(N_{\rm RF}=4\) or \(8\) RF chains, and \(P_{\max}=30\)~dBm. 
Let \(\check{\mathbf r}_{i,j}\) denote the center of $\mathcal C_{i,j}$, where \(i,j\in\{1,2,3,4\}\). Hence, by appropriately mapping $\mathcal C_{i,j}\leftarrow\mathcal C_{n}$: $ \check{\mathbf r}_{i,j}
 =
 \begin{bmatrix}
 \left(i-\frac{5}{2}\right)A\\
 \left(j-\frac{5}{2}\right)A
 \end{bmatrix},
 ~
 \mathcal C_{i,j}
 =
 \check{\mathbf r}_{i,j}
 +
 \left[-\frac{A}{2},\frac{A}{2}\right]^2$, where \(A=\frac{\lambda}{1.5}\). Hence, $\{ \mathcal C_{i,j}\}$ are contiguously tiled without gaps, and their union forms the overall movement aperture $ \mathcal C=\bigcup_{i=1}^{4}\bigcup_{j=1}^{4}\mathcal C_{i,j}$. We set \(d_{\min}=\frac{\lambda}{2}\), $\{\mathbf r_{i,j}^0\}$ are set to the cell centers, i.e., \(\mathbf r_{i,j}^{0}=\check{\mathbf r}_{i,j}\), and $\forall \mathbf z_n^{[0]}=\mathbf e_1$. For channel, the BS-user distance follows i.i.d. $\forall d_k\sim \mathcal U[20, 100]~{\rm m}$, while $\forall\phi_{k,l}\sim \mathcal U\left[-\frac{\pi}{2}, \frac{\pi}{2}\right]$, $\forall\theta_{k,l}\sim \mathcal U[0, \pi]$, \(L=10\), and $T_c=0.25$~s. The path coefficients follow \(\alpha_{k,\ell}\sim\mathcal{CN}(0,1)\) and $ \beta_k=\beta_0 d_k^{-\alpha_{\rm PL}}$ with $\beta_0=-43.45$~dB and \(\alpha_{\rm PL}=2.8\).

For pattern reconfiguration, we employ the $S=8$ distinct radiation patterns from PRA in~\cite{tapbeamp} as in~\cite{ttbd}.\footnote{The EM-domain precoder $\mathbf F_{\rm EM}$ in~\cite{ttbd} is factorized as $\mathbf F_{\rm EM}=(\mathbf I_{N_t}\otimes\mathbf F_{\rm pat})\mathbf S_{\rm EM}$, where $\mathbf S_{\rm EM}=\operatorname{blkdiag}\{\mathbf s_{{\rm EM},1},\cdots,\mathbf s_{{\rm EM},N_t}\}$ and each $\mathbf s_{{\rm EM},n}$ is one-hot. Under $N_t=N$, $P=S$, and $\mathbf s_{{\rm EM},n}=\mathbf z_n$, $\mathbf S_{\rm EM}$ is algebraically identical to $\mathbf F_{\rm sel}(\mathbf Z)$. The difference is that our angle-dependent pattern dictionary is absorbed into $\mathbf H_{\rm sel}(\mathbf R)$, yielding $\mathbf H_{\rm sel}(\mathbf R)\mathbf F_{\rm sel}(\mathbf Z)$, whereas~\cite{ttbd} retains the dictionary explicitly in $\mathbf F_{\rm EM}$; thus, the two formulations are equivalent with respect to discrete per-antenna pattern selection.} 
For position reconfiguration, each MPRA is mounted on a biaxial positioning stage driven by two identical AM2224 stepper motors, with parameters in~\cite[Section~II.B]{Wei2026Mechanical}. Thereby, the maximum axis speed $v_{n,a}^{\max}$ is given by 2.76~m/s. 
Motivated by the compact piezoelectric micropump discussed in~\cite[Section~VI-A]{Wang2025ERFAS}, we adopt a pump power of $P_{\rm pump}=50$~mW and a maximum flow rate of $Q_{\rm pump}=14$~mL/min. Assuming an effective fluid volume of $V_{\rm sw}=0.2~\mu$L per pattern transition, $ \tau_{\rm sw}$ and $ \varepsilon_{\rm sw}$ are respectively set as $
 \tau_{\rm sw}
 =
 \frac{V_{\rm sw}}{Q_{\rm pump}}
 \simeq0.86~{\rm ms}$ and $ \varepsilon_{\rm sw}
 =
 P_{\rm pump}\tau_{\rm sw}
 \simeq42.9~\mu{\rm J}$. The power-related parameters follow~\cite[Table~I and Equation~(13)]{madc}, and all results are averaged over 1,000 Monte Carlo realizations, with every convergence tolerance of $10^{-4}$, which satisfies $ T_c-\epsilon_T>\tau_{\rm sw}$.


For baselines, we compare the proposed \textbf{tri-hybrid MPRA (TH-MPRA)} framework with the following schemes:
\begin{enumerate}

 \item \textbf{Partially-connected hybrid beamforming MA scheme (HBF-MA)~\cite{Wei2026Mechanical}:}
$\mathbf R$ and $(\mathbf F_{\rm BB}, \mathbf F_{\rm RF})$ are optimized, assuming that no pattern-reconfigurable ability on FPA, inspired by~\cite{Wei2026Mechanical}. This incurs mechanical movement time and energy but no pattern-switching overhead.

 \item \textbf{Tri-hybrid pattern-reconfigurable-antenna scheme (TH-PRA)~\cite{Zheng2025TriHybrid}:}
 $\mathbf Z$ (hence $ \mathbf F_{\rm sel}(\mathbf Z)$) and $(\mathbf F_{\rm BB}, \mathbf F_{\rm RF})$ are optimized, while $\mathbf R$ is fixed as $\mathbf R^0$. This incurs pattern-switching overhead but no movement cost.

 \item \textbf{Tri-hybrid beamforming with FPA (TH-FPA)~\cite{trimimo}:}
 $\mathbf R$ is fixed as $\mathbf R^0$, $(\mathbf F_{\rm BB}, \mathbf F_{\rm RF})$ and corresponding EM-domain precoder $\tilde{\mathbf F}_{\rm EM}$ are jointly optimized by~\cite{trimimo}, with no pattern-reconfigurable ability on FPAs on BS. 
 \end{enumerate}

\subsection{Reliability and Optimality of the Proposed Framework}
\label{subsec:reliability_optimality}

\begin{figure}[t]
    \centering
    \subfloat[Initial (open) and optimized (filled) $\mathbf R$ (dashed line: $d_{\rm min}$-radius range).\label{fig:optimized_positions}]{%
        \includegraphics[width=0.25\textwidth]{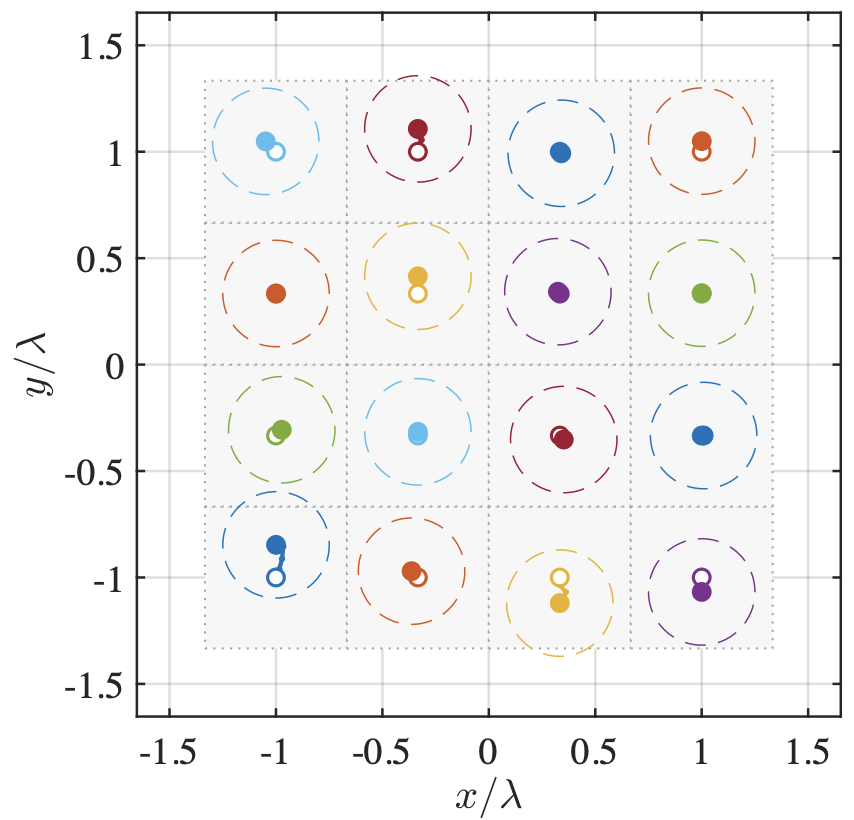}}
\vfill
    \subfloat[Given $\{\theta_{k,l}, \phi_{k,l}\}$ and optimized $\overline{\psi}_n (\boldsymbol{\Omega})$ with $L=5$ and $K=3$.\label{fig:optimized_pattern}]{%
        \includegraphics[width=0.48\textwidth]{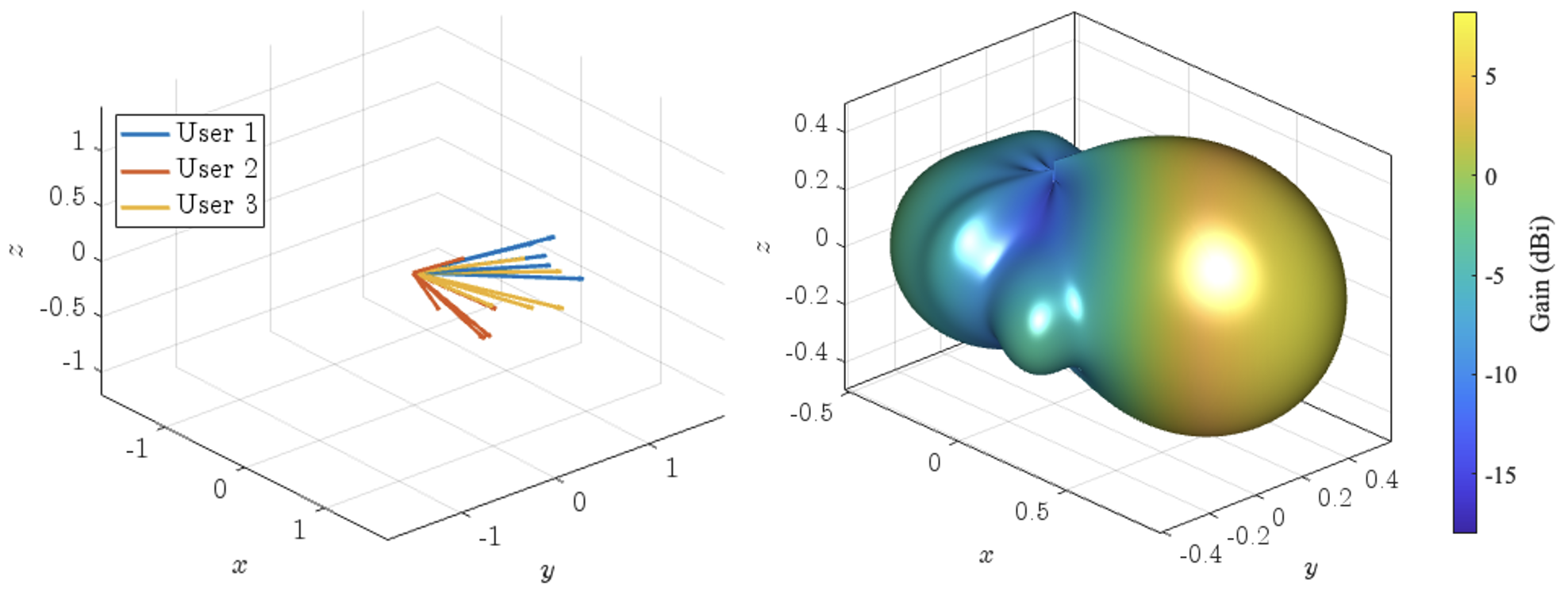}}
    \caption{Representative configurations optimized by the proposed framework.}
    \label{fig:optimized_physical_solution}
\end{figure}
Fig.~\ref{fig:optimized_physical_solution} provides a physical interpretation of optimized configurations. In Fig.~\ref{fig:optimized_positions}, every optimized $\mathbf R$ remains inside its assigned $\{\mathcal C_n\}$, while preserving $d_{\min}$-separation. The MPRA therefore exploits the spatial DoF by proposed framework without producing collision violations. Fig.~\ref{fig:optimized_pattern} further demonstrates how optimizing \(\mathbf Z\) works synergistically with channel environment. The left-panel depicts given $\{\theta_{k,l}, \phi_{k,l}\}$, and the right-panel shows a corresponding $\overline{\psi}_n (\boldsymbol{\Omega})$ selected by the proposed framework~\cite{tapbeamp, ttbd}. The selected $\overline{\psi}_n (\boldsymbol{\Omega})$ redistributes the radiated gain toward favorable angular regions while attenuating less useful directions. Overall, these validate the physical feasibility and effectiveness of the proposed framework, which returns reasonable results thereafter.

\begin{figure}[t]
    \centering
    \subfloat[$N=3$, $N_{\rm RF}=3$, $K=2$, and $L=4$.\label{fig:cdf_case1}]{%
        \includegraphics[width=0.24\textwidth]{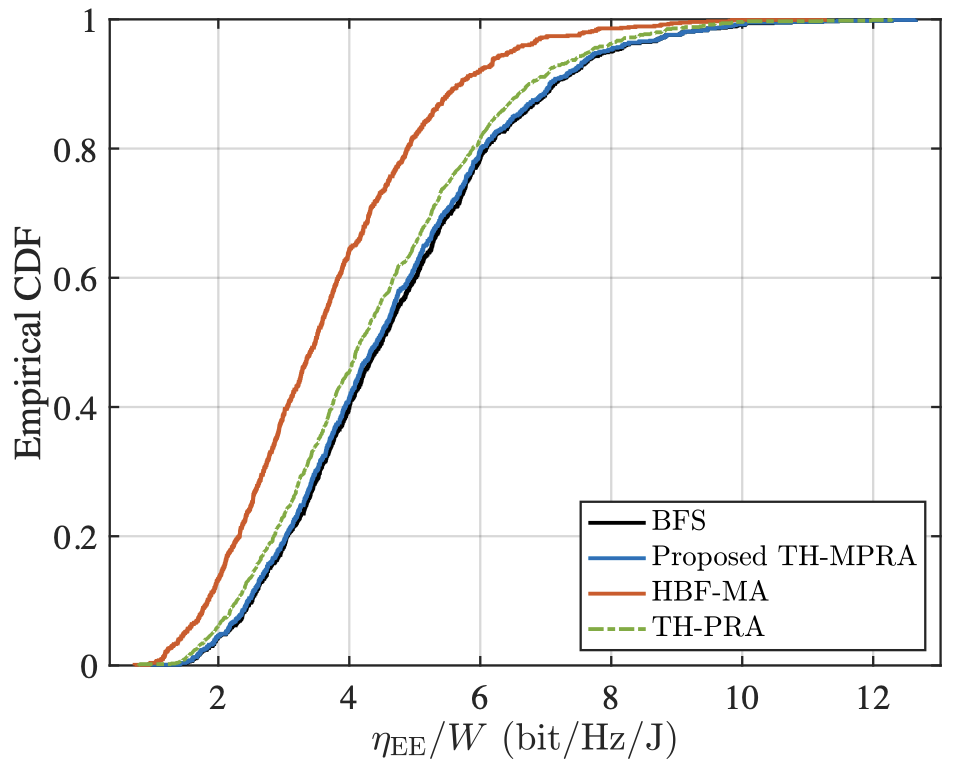}}
    \subfloat[$N=4$, $N_{\rm RF}=2$, $K=2$, and $L=6$.\label{fig:cdf_case2}]{%
        \includegraphics[width=0.24\textwidth]{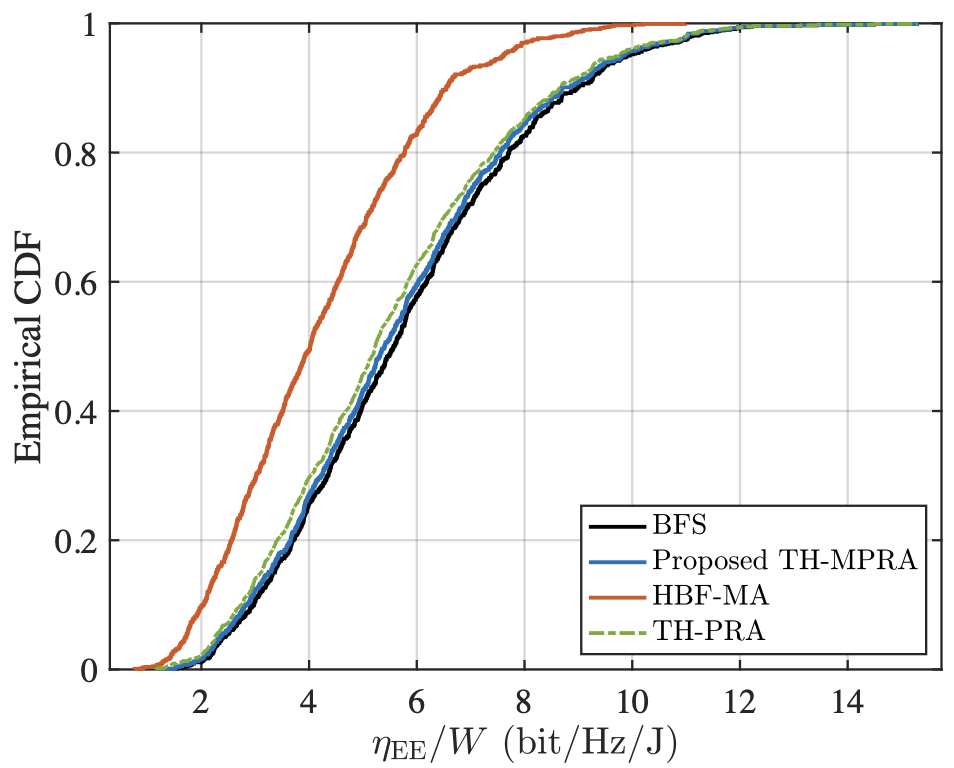}}
    \caption{Empirical CDFs of $\eta_{\rm EE}$ achieved by the BFS reference, the proposed TH-MPRA framework, and baselines.}
    \label{fig:optimality_cdf}
\end{figure}

Fig.~\ref{fig:optimality_cdf} evaluates the optimality of the proposed TH-MPRA framework against BFS, which exhaustively searches all configurations with the continuous variables discretized over a fine grid, for two tractable instances. As shown, the cumulative distribution function (CDF) of the proposed TH-MPRA framework almost overlaps with that of the BFS, demonstrating that it attains near-optimal EE with substantially lower complexity compared to BFS.\footnote{As discussed in Remark~\ref{rem:continuous_position}, this observation further suggests that a discretized implementation would retain performance comparable to the proposed continuous-position framework, with its performance expected to lie within the small gap between the proposed solution and the BFS benchmark.} In contrast, the HBF-MA curve is shifted markedly to the left since it cannot exploit pattern-domain reconfiguration. TH-PRA closes a considerable portion of this gap, while the remaining separation from the proposed framework confirms the additional benefit of optimizing $\mathbf R$.

\begin{figure*}[t]
    \centering
    \subfloat[$(\mathbf F_{\rm BB},\mathbf F_{\rm RF})$-update.\label{fig:conv_f}]{%
        \includegraphics[width=0.24\textwidth]{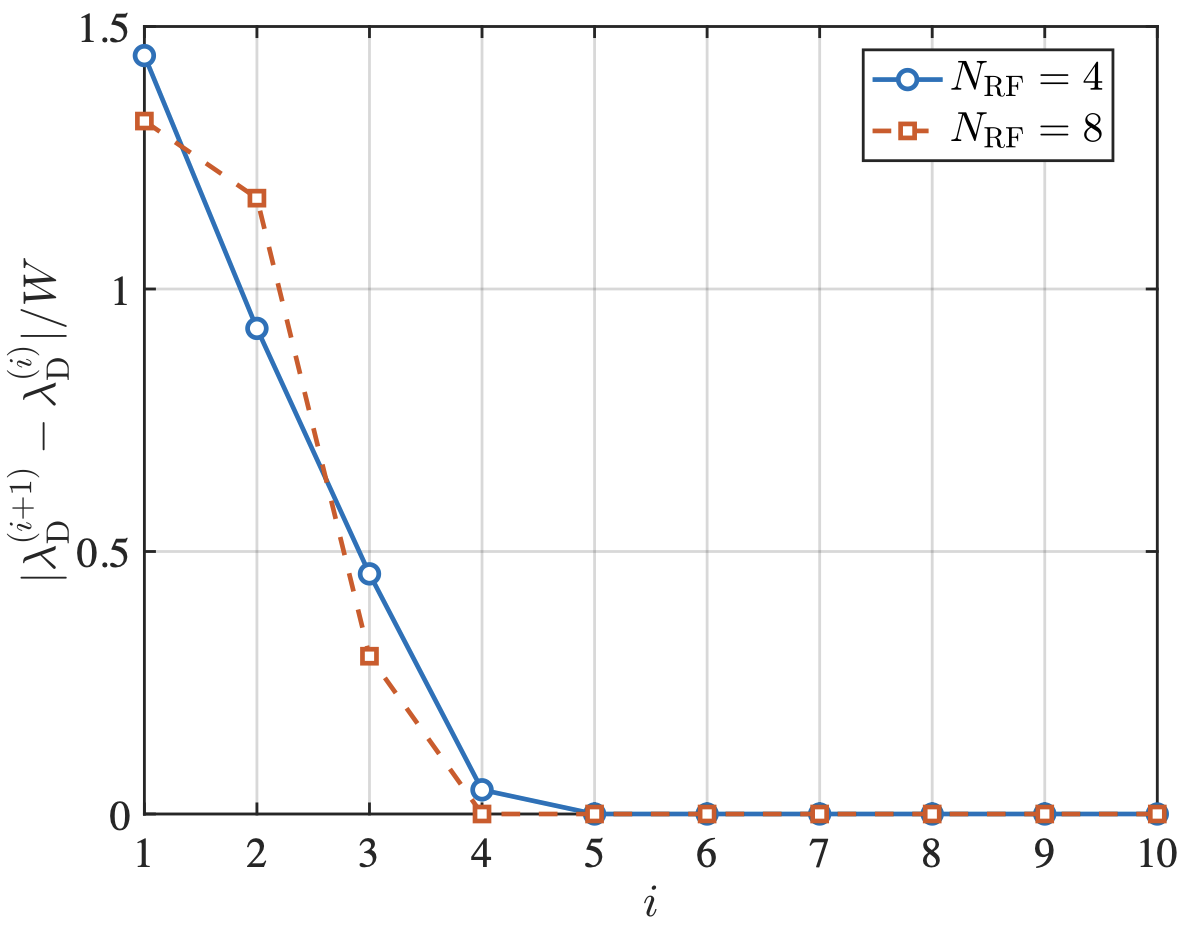}}
    \subfloat[$\mathbf R$-update.\label{fig:conv_r}]{%
        \includegraphics[width=0.24\textwidth]{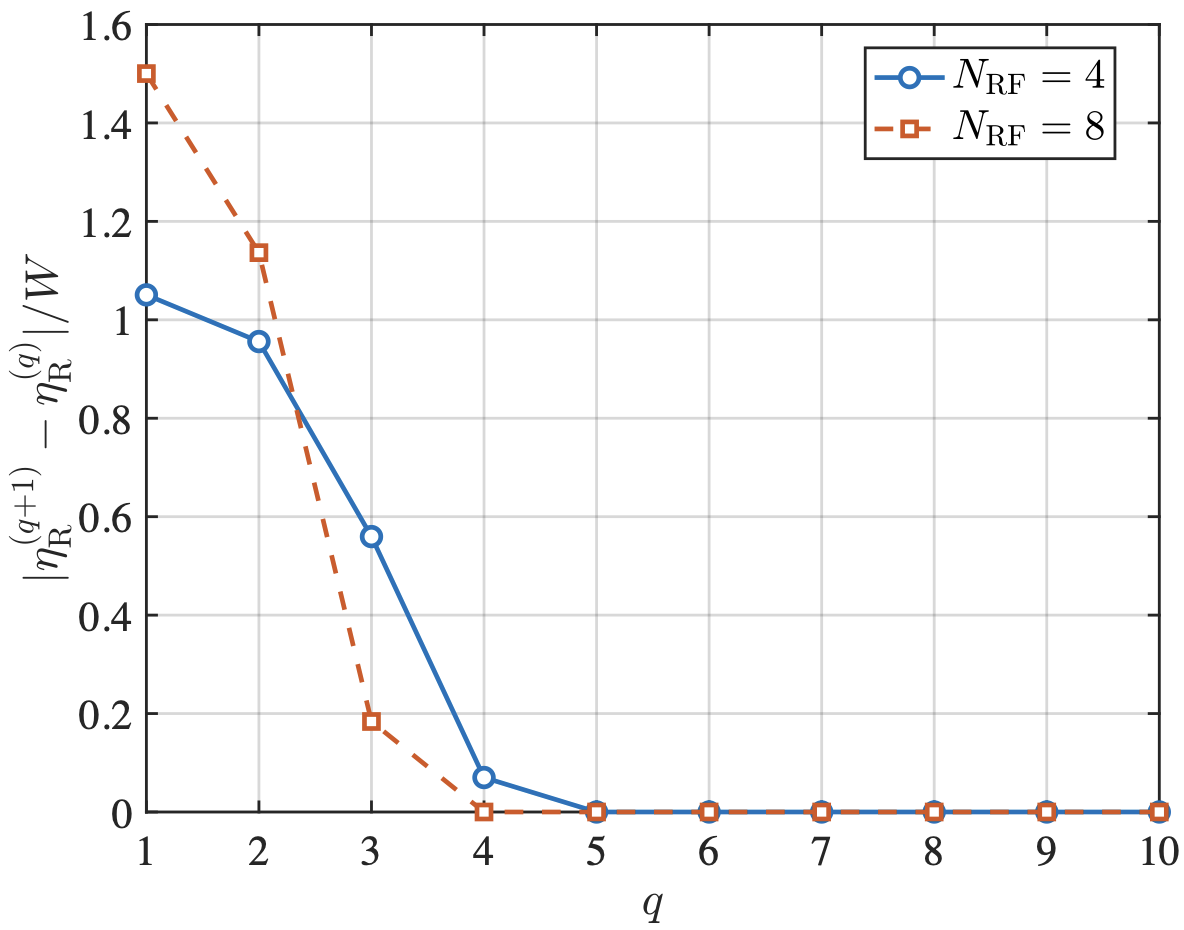}}
    \subfloat[$\mathbf Z$-update.\label{fig:conv_z}]{%
        \includegraphics[width=0.24\textwidth]{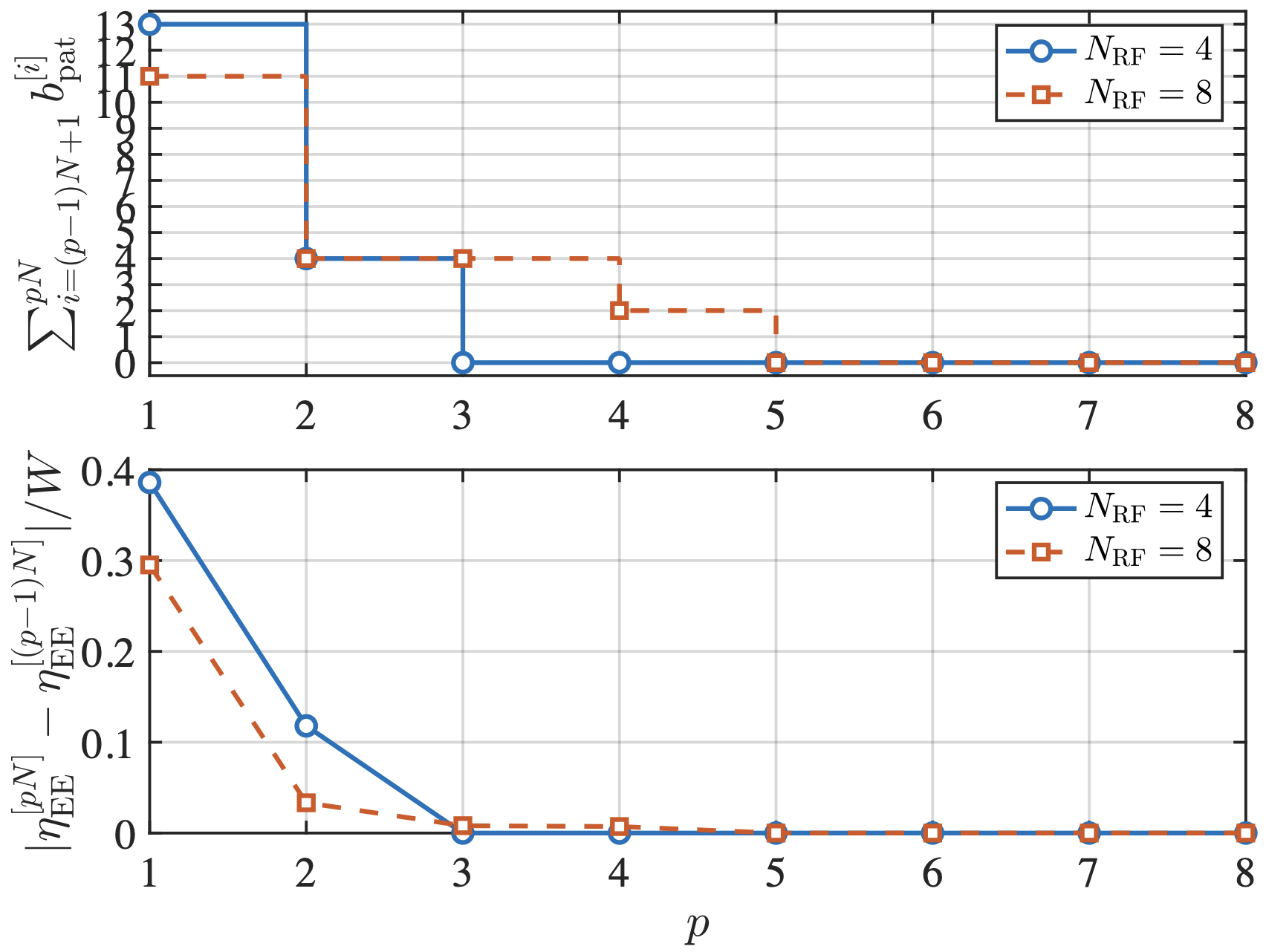}}
    \subfloat[Outer AO framework. \label{fig:conv_ao}]{%
        \includegraphics[width=0.24\textwidth]{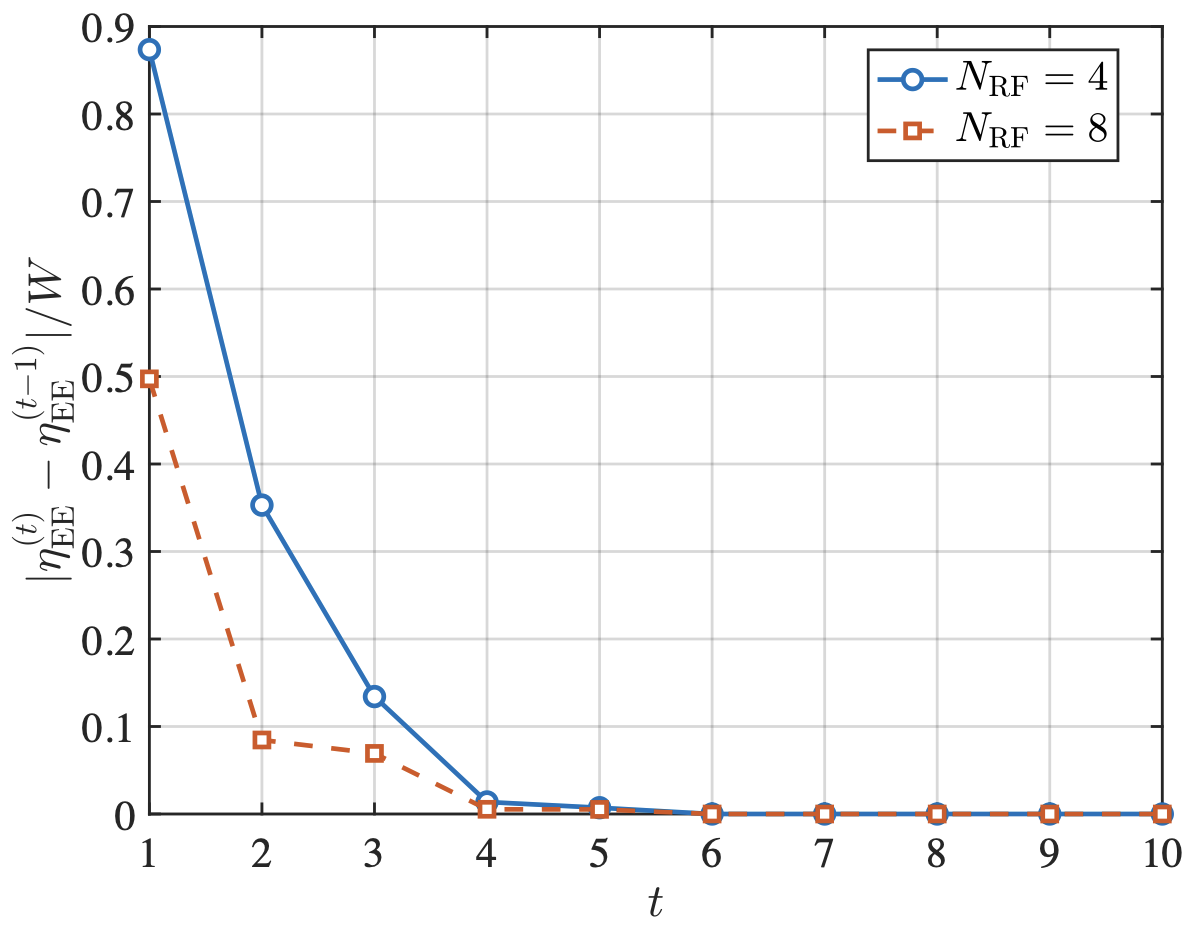}}
    \caption{Convergence behavior of the proposed framework.}
    \label{fig:algorithm_convergence}
\end{figure*}

Fig.~\ref{fig:algorithm_convergence} verifies the convergence reliability of the proposed framework.  Fig.~\ref{fig:conv_f} plots the successive change of $|\lambda_{\rm D}^{(i+1)}-\lambda_{\rm D}^{(i)}|/W$, which rapidly vanishes as $(\mathbf F_{\rm BB}, \mathbf F_{\rm RF})$-subproblem approaches its stationary solution. Fig.~\ref{fig:conv_r} shows a similar decay in $|\eta_{\rm R}^{(q+1)}-\eta_{\rm R}^{(q)}|/W$, confirming stable convergence of the $\mathbf R$-subproblem. For $\mathbf Z$-subproblem, the upper-panel of Fig.~\ref{fig:conv_z} reports the number of antenna-state changes during each complete sweep over $N$ MPRAs ($(p-1)N\sim pN$). The lower panel shows the corresponding sweep-wise variation in the normalized EE. Specifically, both \(\sum_{i=(q-1)N+1}^{qN} b_{\rm pat}^{[i]}\) and \(\lvert\eta_{\rm EE}^{[qN]}-\eta_{\rm EE}^{[(q-1)N]}\rvert/W\) progressively approach zero, indicating that the cyclic pattern search reaches a coordinate-wise stationary point. Finally, Fig.~\ref{fig:conv_ao} also shows that $|\eta_{\rm EE}^{(t)}-\eta_{\rm EE}^{(t-1)}|/W$ decreases to zero within only a few iterations. Overall, the stable convergence behavior clearly demonstrates the reliability.

\begin{figure}[t]
    \centering
    \subfloat[$N_{\rm RF}=4$]{%
        \includegraphics[width=0.24\textwidth]
        {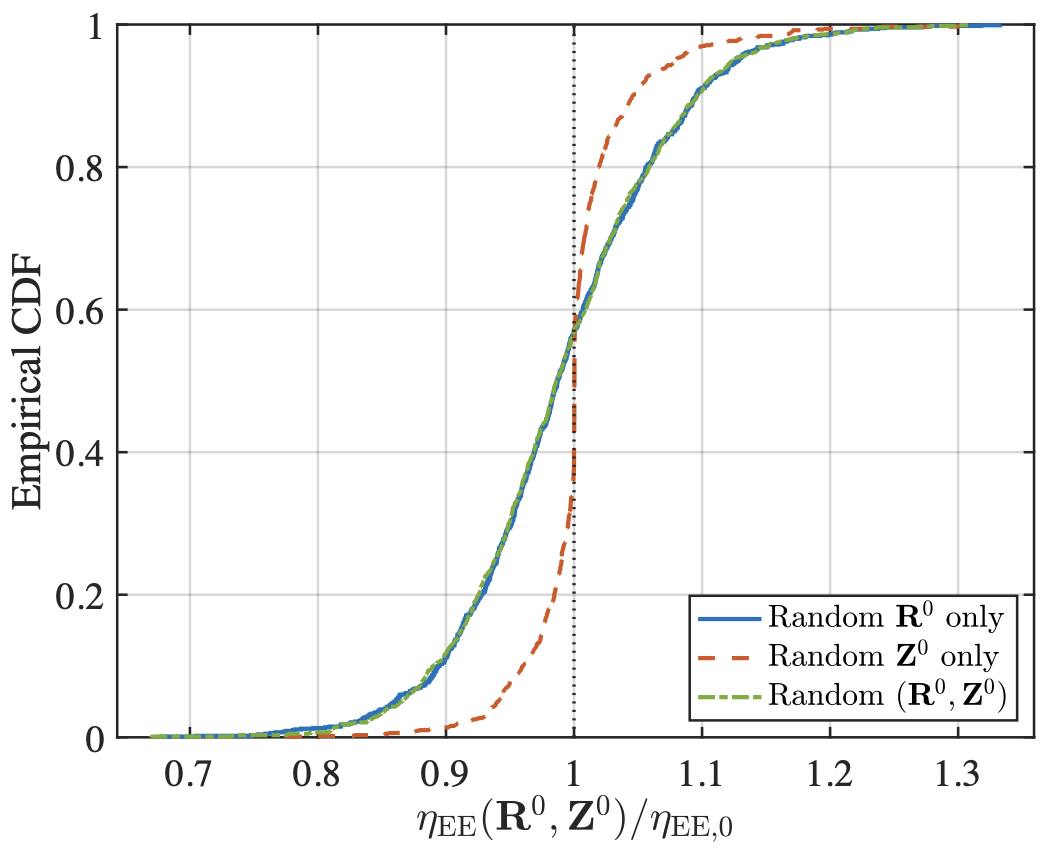}\label{fig:initial_sensitivity_nrf4}}
    \subfloat[$N_{\rm RF}=8$]{%
        \includegraphics[width=0.24\textwidth]
        {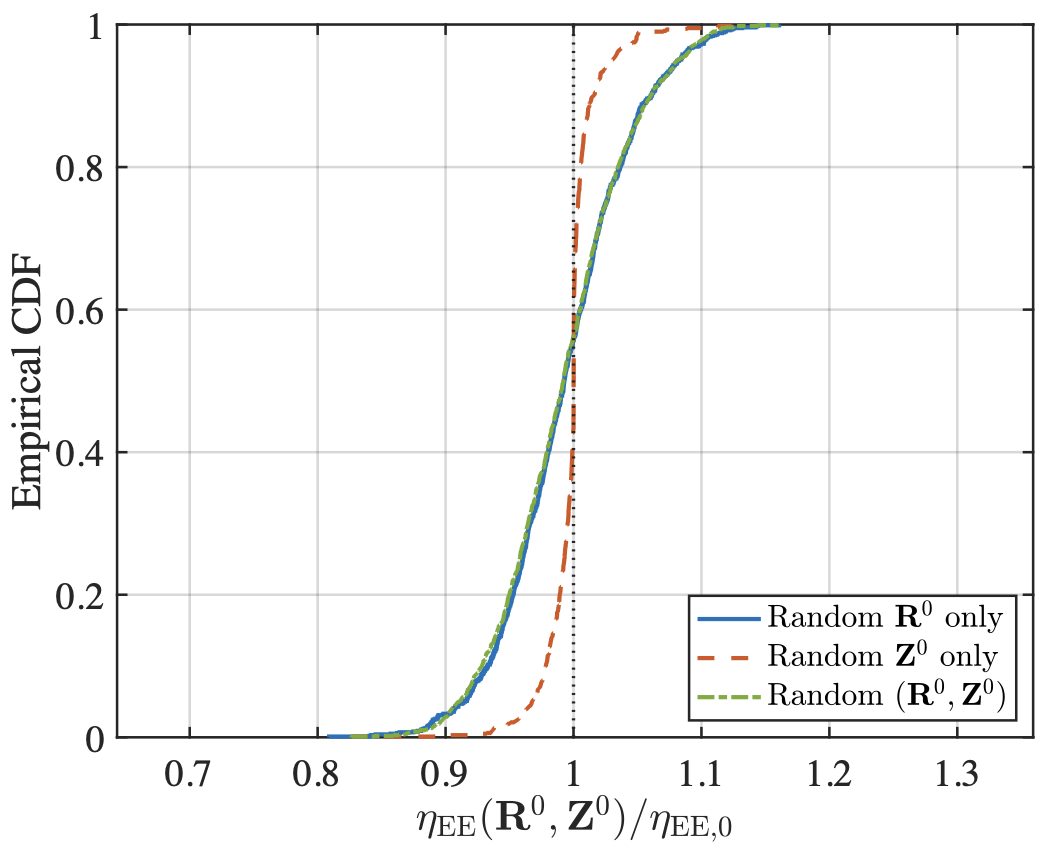}\label{fig:initial_sensitivity_nrf8}}
    \caption{Empirical CDFs of the EE sensitivity under random $(\mathbf R^0, \mathbf Z^0)$.}
    \label{fig:initial_configuration_sensitivity}
\end{figure}

Fig.~\ref{fig:initial_configuration_sensitivity} evaluates initialization sensitivity using \(\rho_{\rm EE}\triangleq\eta_{\rm EE}(\mathbf R^0,\mathbf Z^0)/\eta_{\rm EE,0}\), where \(\eta_{\rm EE,0}\) corresponds to the reference initialization in Section~\ref{subsec:simulation_setup}. Herein, the CDFs concentrate around \(\rho_{\rm EE}=1\); provides representative, although not uniformly optimal, performance. Randomizing only \(\mathbf Z^0\) causes the smallest variation since pattern switching incurs relatively lower overhead and the subsequent \(\mathbf Z\)-update effectively corrects unfavorable states. Conversely, randomizing only \(\mathbf R^0\) produces a wider distribution by altering both channel and movement cost. Its close overlap with the jointly randomized case confirms that sensitivity is dominated by \(\mathbf R^0\). The narrower distributions for \(N_{\rm RF}=8\) further indicate that additional RF-domain flexibility mitigates unfavorable initialization, though its higher $P_{\rm circ}$ does not improve EE.


\subsection{Performance Comparisons Under Various Scenarios}
\label{subsec:performance_comparisons}

\begin{figure*}[t]
    \centering
        \subfloat[]{%
        \includegraphics[width=0.24\textwidth]{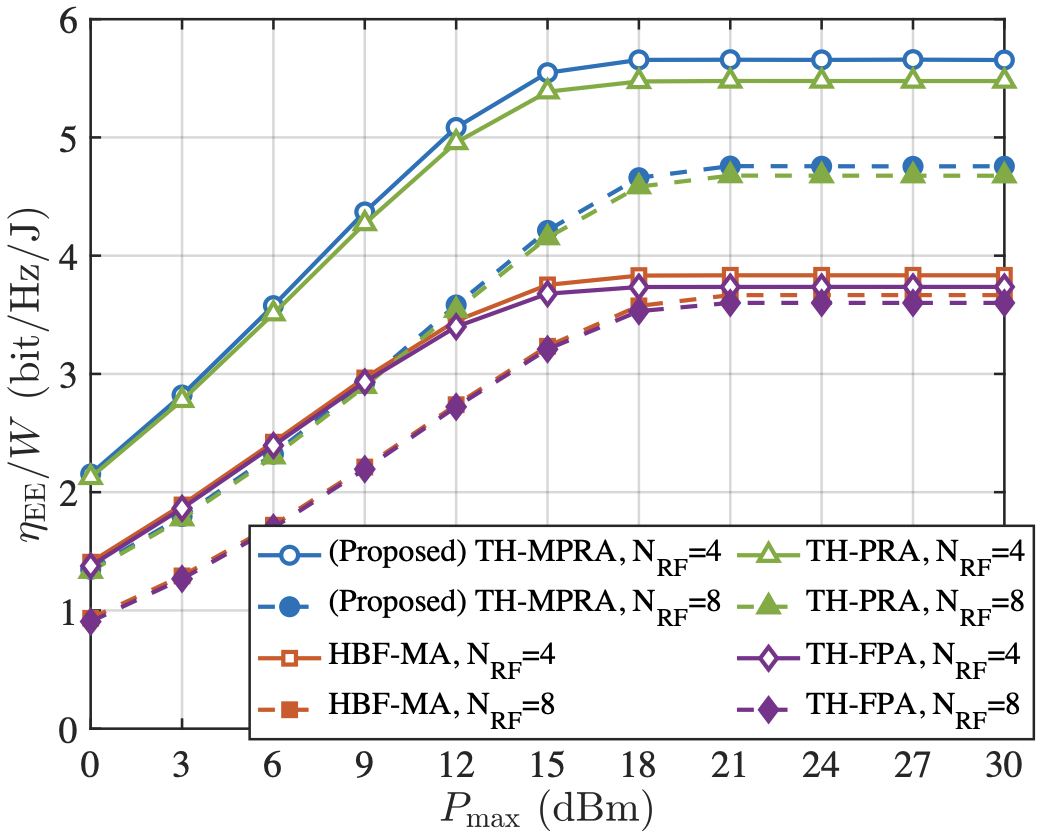} \label{fig:ee_vs_pmax}}
                    \subfloat[]{%
                    \includegraphics[width=0.24\textwidth]{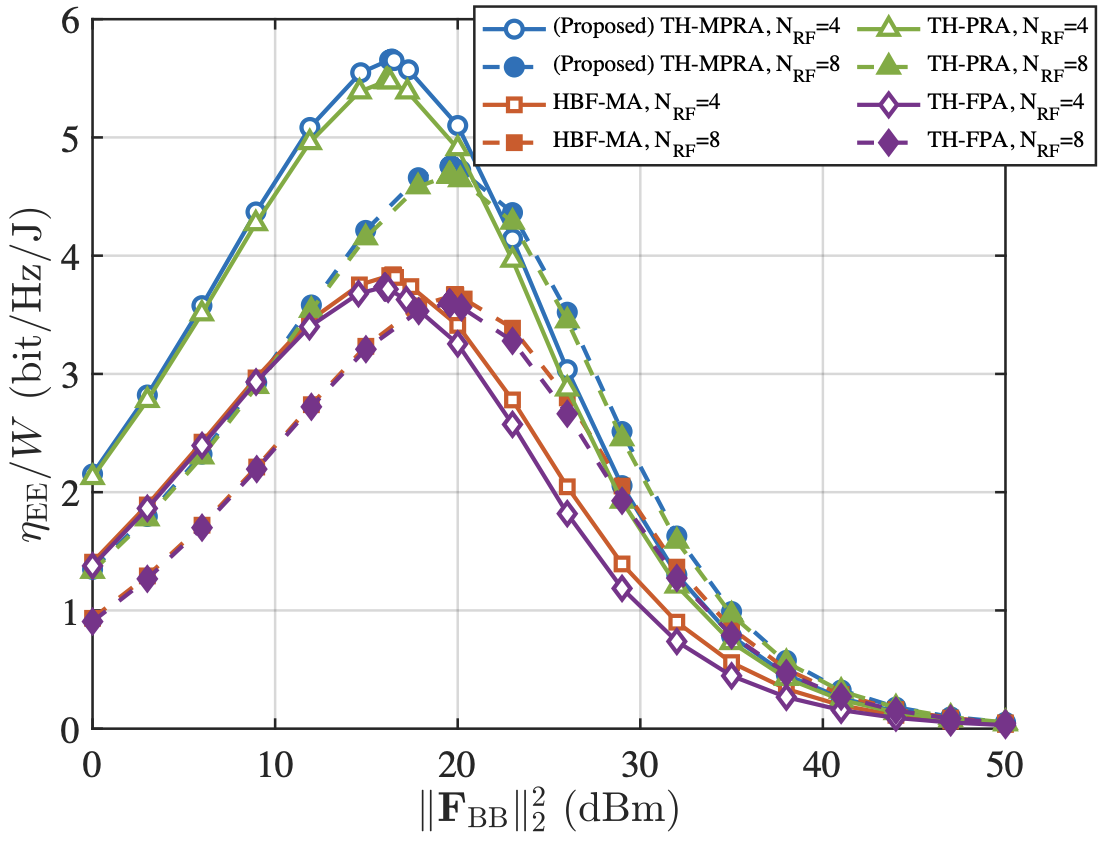} \label{fig:ee_vs_actual_power}}
                            \subfloat[]{%
        \includegraphics[width=0.24\textwidth]{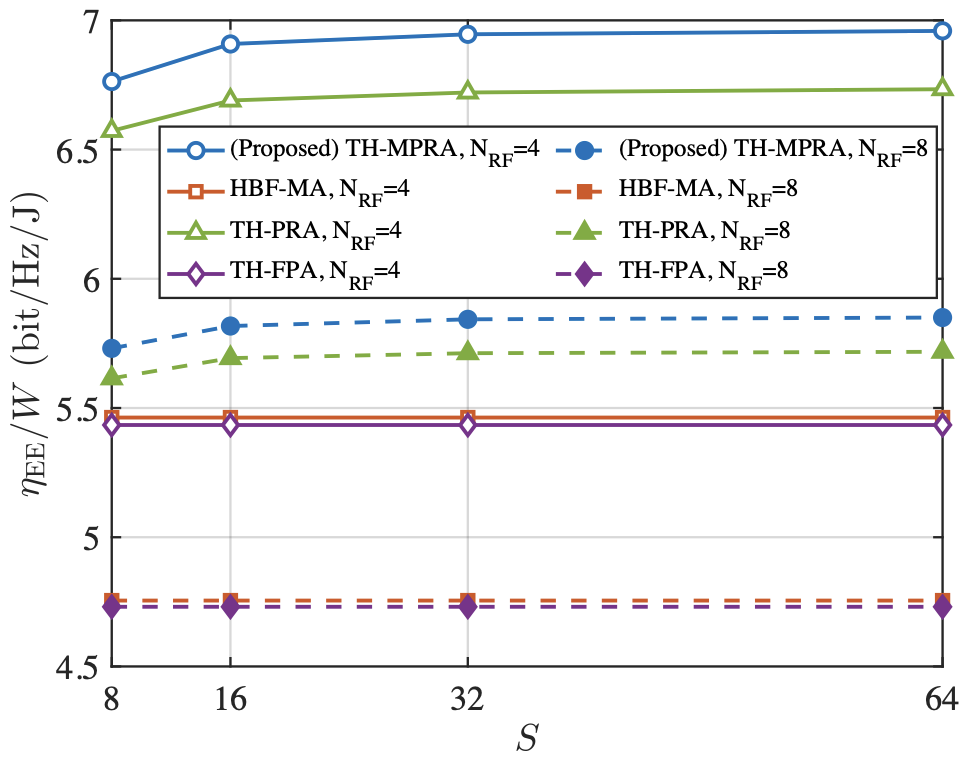} \label{fig:ee_vs_num_patterns}}
                    \subfloat[]{%
                    \includegraphics[width=0.24\textwidth]{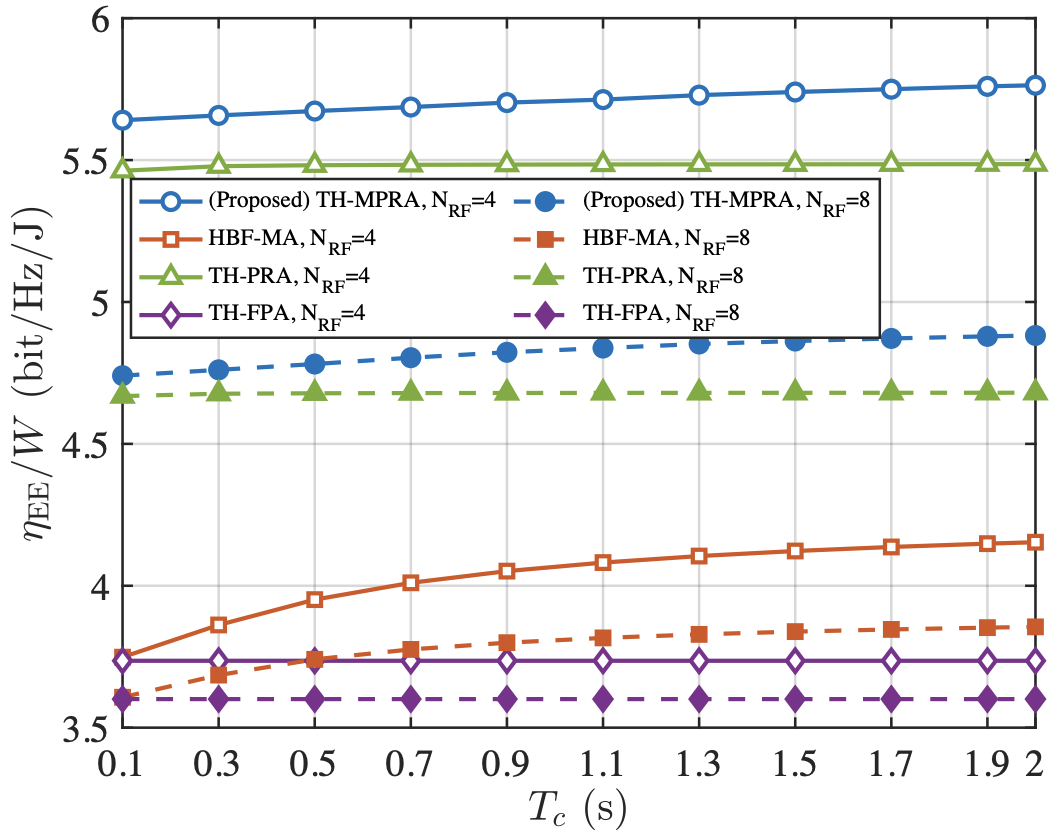} \label{fig:ee_vs_coherence_time}}
    \caption{$\eta_{\rm EE}/W$ versus (a) $P_{\max}$ (b) the actual transmit power $\|\mathbf F_{\rm BB}\|_2$ (c) $S$ and (d) $T_c$.}
    \label{fig:ee_vs_p}
\end{figure*}

\begin{figure*}[t]
    \centering
        \subfloat[]{%
        \includegraphics[width=0.24\textwidth]{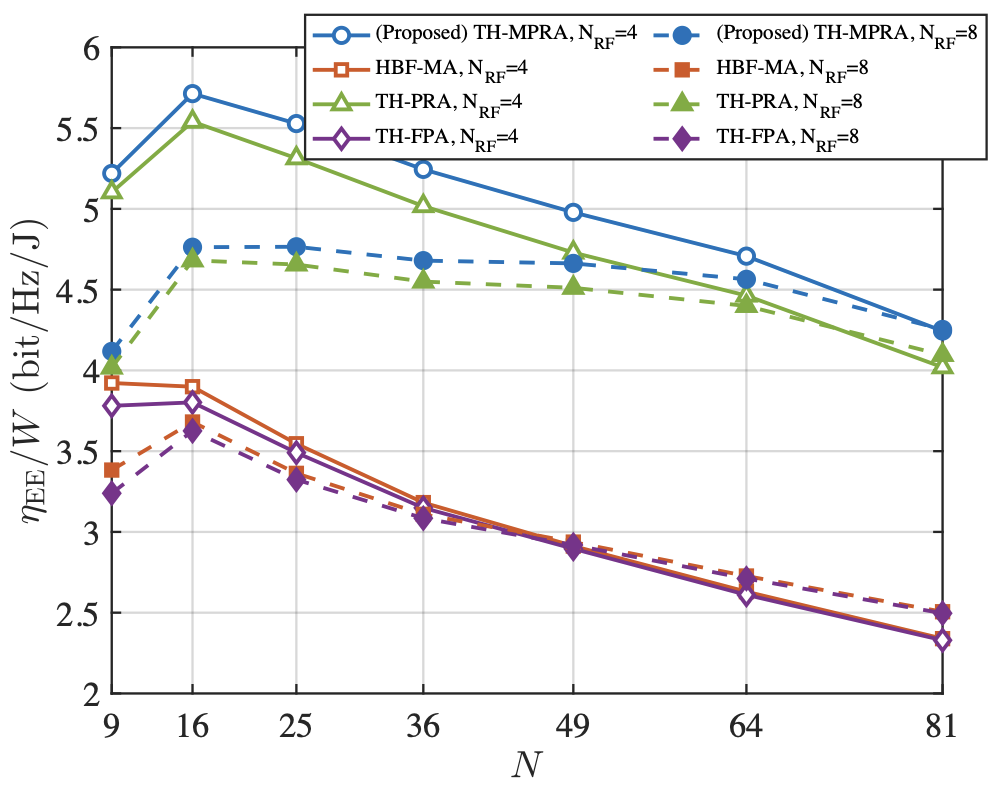} \label{fig:ee_vs_N}}
                    \subfloat[]{%
                    \includegraphics[width=0.24\textwidth]{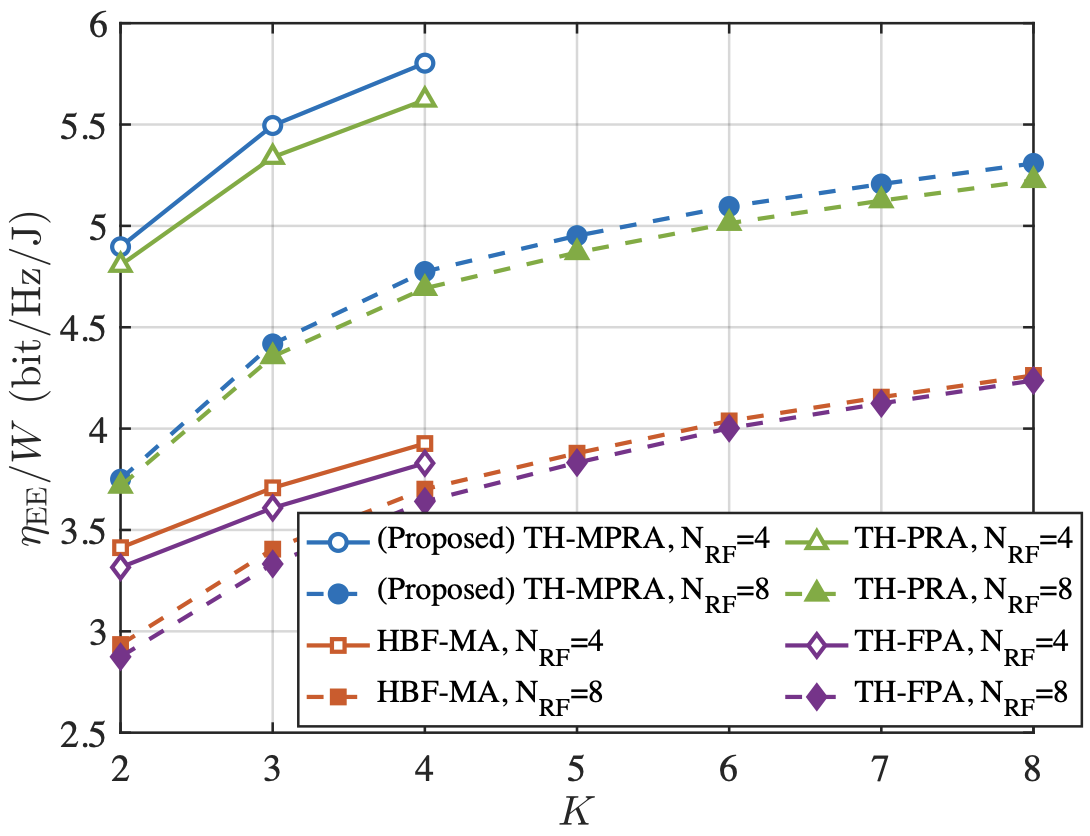} \label{fig:ee_vs_K}}
                            \subfloat[]{%
        \includegraphics[width=0.24\textwidth]{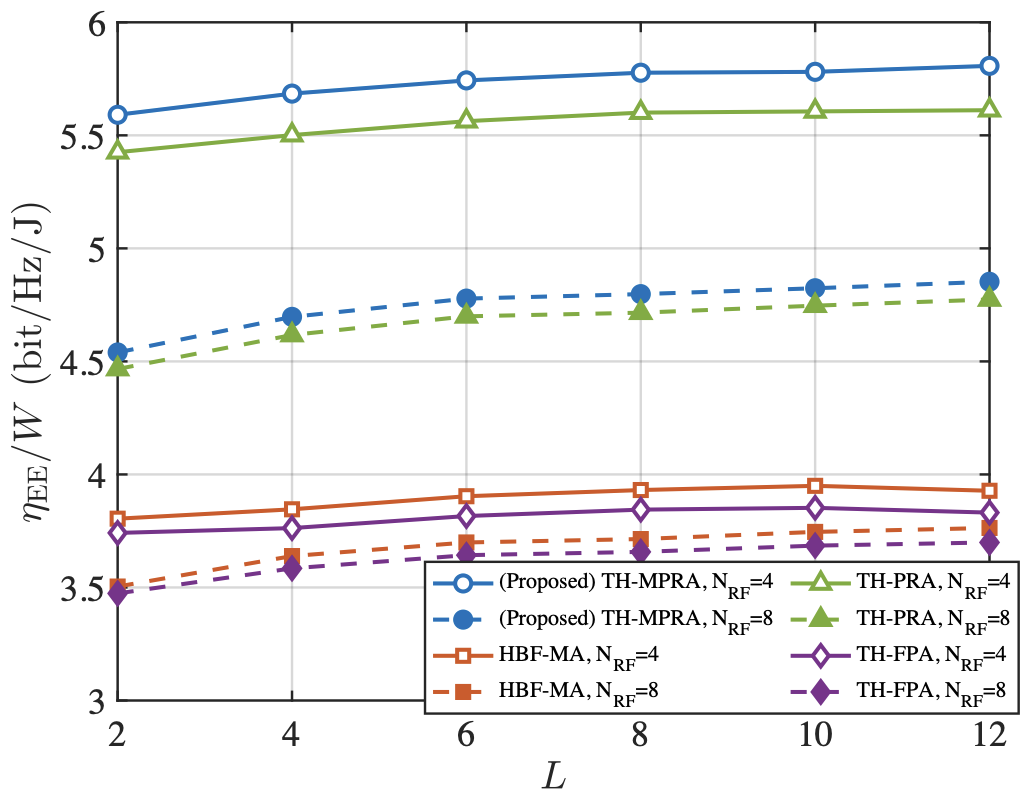} \label{fig:ee_vs_L}}
                    \subfloat[]{%
                    \includegraphics[width=0.24\textwidth]{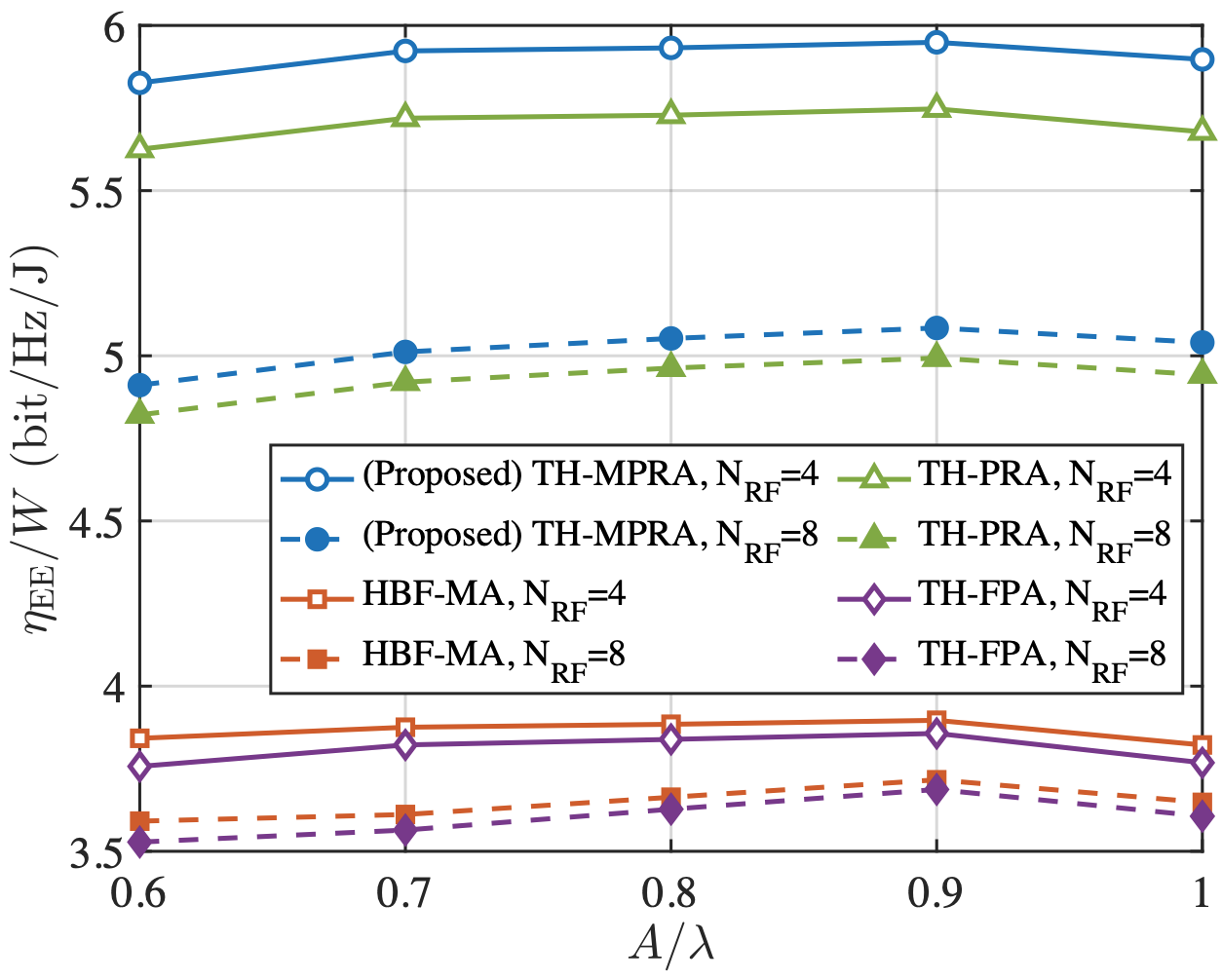} \label{fig:ee_vs_normalized_aperture}}
    \caption{$\eta_{\rm EE}/W$ versus (a) $N$ and (b) $K$ (c) $L$ and (d) $A/\lambda$.}
    \label{fig:ee_vs_nk}
\end{figure*}

Fig.~\ref{fig:ee_vs_pmax} compares the bandwidth-normalized EE, \(\eta_{\rm EE}/W\) versus \(P_{\max}\). As shown, \(\eta_{\rm EE}/W\) initially increases since the SE gain outweighs the additional PA consumption, but saturates since the EE-optimal transmit power is reached and \(\|\mathbf F_{\rm BB}\|_F^2\leq P_{\max}\) becomes inactive. The proposed TH-MPRA framework consistently achieves the highest \(\eta_{\rm EE}/W\); its gain over TH-PRA demonstrates the net benefit of optimizing $\mathbf R$, even after accounting for $E_{\rm mov}$. The observed improvement therefore indicates that the additional spatial DoF can be effectively exploited to improve \(\eta_{\rm EE}/W\) despite the associated movement energy overhead. Meanwhile, the larger separation from HBF-MA and TH-FPA further highlights the effectiveness of adapting $\mathbf Z$. In addition, \(N_{\rm RF}=4\) outperforms \(N_{\rm RF}=8\) since the additional DAC and RF-chain power consumption exceeds the corresponding SE gain, demonstrating that greater RF-domain flexibility does not necessarily improve \(\eta_{\rm EE}/W\).



Fig.~\ref{fig:ee_vs_actual_power} further reveals the EE-power tradeoff by plotting $\eta_{\rm EE}/W$ against the actual transmit power $\|\mathbf F_{\rm BB}\|_F^2$. Unlike Fig.~\ref{fig:ee_vs_pmax}, each curve is unimodal; $\eta_{\rm EE}/W$ first increases since additional $\|\mathbf F_{\rm BB}\|_F^2$ substantially improves the achievable SEs, reaches its maximum at an intermediate power level, and then decreases as the PA energy grows approximately linearly whereas the SE increases only logarithmically. The peak for $N_{\rm RF}=8$ occurs at a slightly higher $\|\mathbf F_{\rm BB}\|_F^2$ than $N_{\rm RF}=4$, since a larger radiated power is required to amortize the higher fixed circuit-power expenditure of the additional RF chains; nevertheless, its maximum $\eta_{\rm EE}/W$ remains lower.  The proposed TH-MPRA framework retains a clear advantage over the baselines, since $\mathbf Z$-update concentrates the radiated energy along favorable multipath directions, while $\mathbf R$-update provides a further channel-conditioning gain.


Fig.~\ref{fig:ee_vs_num_patterns} illustrates the effect of enlarging $ \overline{\boldsymbol{\psi}}(\boldsymbol{\Omega})$. As shown, $\eta_{\rm EE}/W$ of the proposed TH-MPRA framework increases noticeably when $S$ grows from 8 to 16, since a richer codebook provides finer angular selectivity and increases the probability of finding $\overline{\psi}_s(\boldsymbol{\Omega}) $ that reinforces the favorable multipath components while suppressing interference.  The improvement becomes marginal beyond $S=32$, however, since the dominant angular regions have already been covered~\cite{Zheng2025TriHybrid, Wang2025ERFAS, tapbeamp, ttbd}, leads to a clear diminishing return. By construction, HBF-MA and TH-FPA do not optimize $\mathbf Z$ and therefore remain essentially invariant with $S$. The persistent gap between the proposed TH-MPRA framework and TH-PRA represents the complementary gain obtained from $\mathbf R$-update. 
These observations suggest that even a moderate $S=16$ or $32$ offers an attractive performance-search-complexity tradeoff.


Fig.~\ref{fig:ee_vs_coherence_time} illustrates how \(T_c\) affects the cost effectiveness. Herein, $\eta_{\rm EE}/W$ of the proposed TH-MPRA and HBF-MA increases with \(T_c\) since $T_{\rm mov}$ and $E_{\rm mov}$ are 
amortized over a longer data interval. In contrast, TH-PRA varies only slightly since $T_{\rm pat}$ is much smaller than $T_{\rm mov}$ herein (\(T_{\rm pat}=0.86\)~ms vs. \(T_{\rm mov}\approx10.2\)~ms), whereas TH-FPA remains constant since it incurs no reconfiguration overhead. TH-MPRA nevertheless achieves the highest $\eta_{\rm EE}/W$ even for short $T_c$, indicating that its joint optimization gain compensates for the associated overhead. Moreover, the persistent advantage of \(N_{\rm RF}=4\) shows that increasing \(T_c\) does not offset $P_{\rm circ}$.



Fig.~\ref{fig:ee_vs_N} shows how $N$ affects $\eta_{\rm EE}/W$. When $N$ increases from 9 to 16, the array gain improves the effective channel strength and interference-suppression capability, yielding a clear increase in $\eta_{\rm EE}/W$. Beyond this, however, the achievable-SE gain exhibits diminishing returns, whereas $P_{\rm circ}$ continues to grow with $N$. Consequently, the proposed TH-MPRA framework, though it maintains superior performance over baselines, attains their largest $\eta_{\rm EE}/W$ around $N=16$ and then gradually decreases, and so is TH-PRA. This degradation is more pronounced for HBF-MA and TH-FPA since they cannot use pattern reconfiguration to extract sufficient channel gain from the enlarged $N$. The result therefore reveals an $\eta_{\rm EE}$-optimal $N$ rather than an ``as-large-as-possible'' rule.


Fig.~\ref{fig:ee_vs_K} depicts the effect of $K$.\footnote{For $N_{\rm RF}=4$, only $K\leq4$ is shown due to $K\leq N_{\rm RF}$~\cite{hbyx, madc}.}  As shown, $\eta_{\rm EE}/W$ increases with $K$ since the spatial multiplexing delivers more aggregate information during the same coherence block and amortizes $P_{\rm circ}$ and reconfiguration costs over a larger SE.  The increase becomes progressively smaller at large $K$ owing to stronger multi-user interference, transmit-power sharing, and the finite DoF. The proposed TH-MPRA framework consistently outperforms the benchmarks, with the gap over TH-PRA reflecting the additional channel orthogonalization provided by position control and the larger gaps over HBF-MA and TH-FPA demonstrating the importance of pattern-domain interference management. Over $K=2-4$, $N_{\rm RF}=4$ achieves higher $\eta_{\rm EE}/W$ because of its lower circuit-power expenditure. 


Fig.~\ref{fig:ee_vs_L} demonstrates that richer multipath is beneficial but produces only a moderate improvement on $\eta_{\rm EE}/W$.  As $L$ increases, the additional directions provide greater spatial diversity~\cite{hjdrl, Zheng2025TriHybrid}, allowing TH-MPRA to jointly select favorable $\mathbf R$ and $\mathbf Z$. The gain gradually saturates since the channel model normalizes the path powers with respect to $L$, hence additional paths increase diversity rather than the received power; moreover, newly added paths can be angularly correlated, or interference generating. Nevertheless, the proposed TH-MPRA framework maintains the largest $\eta_{\rm EE}/W$ throughout, with its persistent gains over TH-PRA, HBF-MA, and TH-FPA highlighting the benefits of jointly exploiting reconfiguration of $\mathbf R$ and $\mathbf Z$, which provides a complementary phase-domain DoF even in rich scattering scenarios.


Fig.~\ref{fig:ee_vs_normalized_aperture} illustrates the effect of \(A/\lambda\). Herein, 
the result reflects the combined effects of array geometry and, for frameworks with position adaptation, antenna movement. 
For TH-MPRA, increasing \(A/\lambda\) from 0.6 to 0.8-0.9 improves \(\eta_{\rm EE}/W\), as the expanded array geometry and positioning flexibility can enhance spatial phase diversity and user separation. As \(A/\lambda\) approaches 1, however, it yields a slightly lower EE. Larger positioning regions can also entail longer movement distances, increasing \(T_{\rm mov}\) and \(E_{\rm mov}\), although their contribution cannot be isolated from the accompanying geometry changes in this experiment. Accordingly, \(A/\lambda=\)0.8-0.9 provides the most favorable balance under the considered setup. Throughout the evaluated range, TH-MPRA consistently outperforms the baselines by jointly exploiting position and pattern adaptation.

\section{Conclusion}
\label{sec:conclusion}
In this paper, we proposed an energy-efficient multi-user framework with tri-hybrid MPRA. A hardware-inspired power and latency model captured circuit consumption, mechanical and pattern actuation, and transmission-time overhead. A unified AO framework addressed the coupled EE maximization through block-specific optimization techniques, guaranteeing a nondecreasing and convergent EE sequence. Simulations demonstrated performance close to exhaustive search and consistent EE gains over the baselines. These highlight the potential of joint spatial and pattern reconfiguration on MA for energy-efficient wireless network in 6G and beyond.

\bibliographystyle{IEEEtran}
\bibliography{IEEEexample}

\end{document}